\documentclass{CSML}
\pdfoutput=1

\usepackage{lastpage}
\newcommand{\dOi}{13(4:1)2017}
\lmcsheading{}{1--\pageref{LastPage}}{}{}%
{Dec.~09,~2016}{Oct.~23,~2017}{}

\usepackage{enumerate}
\usepackage{hyperref}
\hypersetup{hidelinks}
\usepackage{cite}
\usepackage{amssymb}
\usepackage{amsmath}
\usepackage{amsthm}
\usepackage{proof}
\usepackage{stmaryrd}
\usepackage{tikz}
\usetikzlibrary{arrows}

\newcommand{\U}{\mathsf{U}}
\newcommand{\El}{\mathsf{El}}
\newcommand{\REN}{\mathsf{REN}}
\newcommand{\op}{\mathsf{op}}
\newcommand{\ra}{\rightarrow}

\newcommand{\Set}{\mathsf{Set}}
\newcommand{\PSh}{\mathsf{PSh}}
\newcommand{\FamPSh}{\mathsf{FamPSh}}
\renewcommand{\ll}{\llbracket}
\providecommand{\rr}{\rrbracket}
\newcommand{\Con}{\mathsf{Con}}
\newcommand{\Ty}{\mathsf{Ty}}
\newcommand{\Tm}{\mathsf{Tm}}
\newcommand{\Tms}{\mathsf{Tms}}
\newcommand{\R}{\mathsf{R}}
\newcommand{\TM}{\mathsf{TM}}
\newcommand{\NE}{\mathsf{NE}}
\newcommand{\NF}{\mathsf{NF}}
\newcommand{\q}{\mathsf{q}}
\renewcommand{\u}{\mathsf{u}}

\newcommand{\cul}{\ulcorner}
\newcommand{\cur}{\urcorner}
\newcommand{\norm}{\mathsf{norm}}
\newcommand{\Nf}{\mathsf{Nf}}
\newcommand{\Ne}{\mathsf{Ne}}

\newcommand{\id}{\mathsf{id}}
\newcommand{\nat}{\,\dot{\rightarrow}\,}

\renewcommand{\S}{\overset{\mathsf{s}}{\ra}} 
\newcommand{\ap}{\mathsf{ap}}
\newcommand{\blank}{\mathord{\hspace{1pt}\text{--}\hspace{1pt}}} 
\newcommand{\lam}{\mathsf{lam}}
\newcommand{\app}{\mathsf{app}}
\newcommand{\circid}{\circ\hspace{-0.2em}\id}
\newcommand{\circcirc}{\circ\hspace{-0.2em}\circ}
\newcommand{\tr}[2]{\ensuremath{{}_{#1 *}\mathopen{}{#2}\mathclose{}}} 
\newcommand{\M}{\mathsf{M}}
\newcommand{\C}{\mathcal{C}}
\newcommand{\data}{\mathsf{data}}
\newcommand{\ind}{\hspace{1em}}
\newcommand{\idP}{\mathsf{idP}}
\newcommand{\compP}{\mathsf{compP}}
\newcommand{\idF}{\mathsf{idF}}
\newcommand{\compF}{\mathsf{compF}}
\newcommand{\proj}{\mathsf{proj}}

\newcommand{\map}{\mathsf{map}}
\newcommand{\Var}{\mathsf{Var}}
\newcommand{\Vars}{\mathsf{Vars}}
\newcommand{\vze}{\mathsf{vze}}
\newcommand{\vsu}{\mathsf{vsu}}
\newcommand{\wk}{\mathsf{wk}}
\newcommand{\wkV}{\mathsf{wkV}}
\newcommand{\neuU}{\mathsf{neuU}}
\newcommand{\neuEl}{\mathsf{neuEl}}
\newcommand{\var}{\mathsf{var}}
\newcommand{\natn}{\mathsf{natn}}
\newcommand{\natS}{\mathsf{natS}}
\newcommand{\LET}{\mathsf{let}}
\newcommand{\IN}{\mathsf{in}}
\newcommand{\refl}{\mathsf{refl}}
\newcommand{\trans}{\mathbin{\raisebox{0.5ex}{$\displaystyle\centerdot$}}}
\renewcommand{\P}{\mathsf{P}}
\renewcommand{\r}[1]{{\P_{#1}}}
\newcommand{\inj}{\mathsf{inj}}

\newcommand{\Elim}{\mathsf{Elim}}
\renewcommand{\tt}{\mathsf{tt}}
\newcommand{\NTy}{\mathsf{NTy}}
\newcommand{\dec}{\mathsf{dec}}
\newcommand{\Dec}{\mathsf{Dec}}
\newcommand{\lb}{\langle}
\newcommand{\rb}{\rangle}

\newcommand{\stab}{\mathsf{stab}}
\newcommand{\cons}{\mathsf{cons}}

\newcommand\arcfrombottom{
  \begin{tikzpicture}[scale=0.03em]
    \draw (0,0) arc (0:180:0.5);
    \draw (0,0) edge[->] (0,-0.3);
    \draw (-1,0) edge (-1,-0.3);
  \end{tikzpicture}
}
\newcommand\arcfromtop{
  \begin{tikzpicture}[scale=0.03em]
    \draw (0,0) arc (180:360:0.5);
    \draw (0,0) edge[->] (0,0.3);
    \draw (1,0) edge (1,0.3);
  \end{tikzpicture}
}

\begin{document}

\title[Normalisation by Evaluation for Type Theory, in Type Theory]{Normalisation by Evaluation for Type Theory, \\in Type Theory}


\author[Altenkirch and Kaposi]{Thorsten Altenkirch}	
\address{School for Computer Science, University of Nottingham, Nottingham, United Kingdom}	
\email{txa@cs.nott.ac.uk}  
\thanks{This research was supported by EPSRC grant EP/M016951/1, USAF grant FA9550-16-1-0029 and COST Action EUTypes CA15123.}	

\author[]{Ambrus Kaposi}	
\address{Department of Programming Languages and Compilers, E{\"o}tv{\"o}s Lor{\'a}nd University, Budapest, Hungary}	
\email{akaposi@inf.elte.hu}  





\begin{abstract}
  \noindent We develop normalisation by evaluation (NBE) for dependent
  types based on presheaf categories. Our construction is formulated
  in the metalanguage of type theory using quotient inductive
  types. We use a typed presentation hence there are no preterms or
  realizers in our construction, and every construction respects the
  conversion relation. NBE for simple types uses a logical relation
  between the syntax and the presheaf interpretation. In our
  construction, we merge the presheaf interpretation and the logical
  relation into a proof-relevant logical predicate. We prove
  normalisation, completeness, stability and decidability of
  definitional equality. Most of the constructions were formalized in
  Agda.
\end{abstract}


\maketitle


\section{Introduction}

Normalisation by evaluation (NBE) is a technique to compute normal forms of
typed $\lambda$-terms by evaluating them in an appropriate
semantics. The idea was pioneered  by Schwichtenberg and
Berger \cite{berger1991inverse}, subsequently a categorical account
using presheaf categories was given \cite{alti:ctcs95} and this
approach was extended to System F \cite{alti:lics96,alti:f97} and
coproducts \cite{alti:lics01}. 

In the present paper we extend NBE to a basic type theory with
dependent types which has $\Pi$-types and an uninterpreted
family using a presheaf interpretation.  We take advantage of our recent work on an intrinsic
representation of type theory in type theory \cite{ttintt} which
only defines typed objects avoiding any reference to untyped preterms
or typing relations and which forms the basis of our formal
development in Agda. 

The present paper is an expanded version of our conference paper
\cite{nbe-conf}. In particular we show here for the first time that
our normalisation construction implies decidability of equality. This
isn't entirely obvious because our normal forms are indexed by
contexts and types of which it is a priori not known wether equality
is decidable. However, we observe that mimicking the bidirectional
approach to type checking \cite{Coquand96analgorithm} we can actually
decide equality of normal forms and hence, after combining it with
normalisation, we obtain decidability for conversion.

\subsection{Specifying normalisation}
\label{sec:spec}

Normalisation can be given the following specification.

We denote the type of well typed terms of type $A$ in context $\Gamma$
by $\Tm \, \Gamma \, A$. We are not interested in preterms, all of our
constructions will be well-typed. In addition, this type is defined as
a quotient inductive type (QIT, see \cite{ttintt}) which means that
terms are quotiented with the conversion relation. It follows that on
one hand if two terms $t, t' : \Tm\,\Gamma\,A$ are convertible then
they are equal: $t \equiv_{\Tm \, \Gamma \, A} t'$. On the other hand,
the eliminator of $\Tm\,\Gamma\,A$ ensures that every function defined
from this type respects the conversion relation. This enforces a high
level of abstraction when reasoning about the syntax: all of our
constructions need to respect convertibility as well.

The type of normal forms is denoted $\Nf \, \Gamma \, A$ and there is
an embedding from it to terms $\cul \blank \cur : \Nf \, \Gamma \, A
\ra \Tm\,\Gamma\,A$. Normal forms are defined as a usual inductive
type (as opposed to quotient inductive types).

Normalisation is given by a function $\norm$ which takes a term to a
normal form. It needs to be an isomorphism:
\begin{equation*}
   \text{completeness }\arcfromtop \hspace{2em} \norm\downarrow\begin{array}{l}\infer={\hspace{1em} \Nf \, \Gamma \, A \hspace{1em} }{\Tm \, \Gamma \, A}\end{array}\uparrow\cul\blank\cur \hspace{2em} \arcfrombottom\text{ stability}
\end{equation*}
If we normalise a term, we obtain a term which is convertible to it:
$t \equiv \cul \norm \, t \cur$. This is called completeness. The
other direction is called stability: $n \equiv \norm \, \cul n
\cur$. It expresses that there is no redundancy in the type of normal
forms. Soundness, that is, if $t \equiv t'$ then $\norm \, t \equiv
\norm \, t'$ is given by congruence of equality.

\subsection{NBE for simple type theory}

Normalisation by evaluation (NBE) is one way to implement this
specification. It works by a complete model construction (figure
\ref{fig:nbe}). We define a model of the syntax and hence the
eliminator gives us a function from the syntax to the model. Then we
define a quote function which is a map from the model back to the
syntax, but it targets normal forms (a subset of the syntax via the
operator $\cul\blank\cur$). \\
\vspace{-1em}
\begin{figure}[h]
\centering
\begin{tikzpicture}
\node (Syntax) at (2,3.6) {Syntax};
\node (Model) at (7,3.5) {Model};
\node (NF) at (2,2) {\begin{tabular}{c} Normal \\ forms \end{tabular}};
\draw (2,2) circle (1.3cm);
\draw (2,2) circle (0.8cm);
\draw (7,2) circle (1.2cm);
\draw[thick,->] (3.3,2.5) -- node[above] {eliminator} (5.8,2.5);
\draw[thick,->] (5.8,1.5) -- node[above] {quote} (2.7,1.5);
\end{tikzpicture}
\caption{Normalisation by evaluation.}
\label{fig:nbe}
\end{figure}

In this subsection, we summarize the approach of \cite{alti:ctcs95}
for NBE for simple types. Here the model we choose is a presheaf
model. Presheaf models are proof-relevant versions of Kripke models
(possible world semantics) for intuitionistic logic: they are
parameterised over a category instead of a poset. The category that we
choose here is the category of renamings $\REN$. The objects in $\REN$
are contexts and morphisms are variable renamings. The presheaf model
interprets contexts and types as presheaves, e.g.\ the interpretation
of $A$ denoted $\ll A\rr$ is a $\REN^\op \ra \Set$ functor. Terms and
substitutions are natural transformations between the corresponding
presheaves, e.g.\ for $t : \Tm\,\Gamma\,A$ we have a natural
transformation $\ll t\rr : \ll\Gamma\rr\nat\ll A\rr$. A function type
is interpreted as the presheaf exponential (a function for all future
worlds), the base type is interpreted as normal forms of the base
type.

Because $\REN$ has contexts as objects, we can embed types into
presheaves (Yoneda embedding): a type $A$ is embedded into the
presheaf $\TM_A$ by setting $\TM_A\,\Psi = \Tm\,\Psi\,A$ i.e.\ a type
at a given context is interpreted as the set of terms of that type in
that context. Analogously, we can embed a type $A$ into the
presheaveas of normal forms $\NF_A$ and neutral terms $\NE_A$. Normal
forms are terms with no redexes (they include neutral terms) while
neutral terms are either variables or an eliminator applied to a
neutral term.

The quote function is defined by induction on types as a natural
transformation $\q_A : \ll A \rr \nat \NF_A$. Quote is defined
mutually with unquote which maps neutral terms into semantic elements:
$\u_A : \NE_A \nat \ll A \rr$.

To normalise a term, we also need to define unquote for neutral
substitutions (lists of neutral terms). Then we get normalisation by
calling unquote on the identity neutral substitution, then
interpreting the term at this semantic element and finally quoting.
\begin{equation*}
  \norm_A\,(t:\Tm\,\Gamma\,A) : \Nf\,\Gamma\,A := \q_{A}\,\big(\ll t\rr\,(\u_\Gamma\,\id)\big)
\end{equation*}

We can prove completeness using a logical relation $\R$ between $\TM$
and the presheaf model. The logical relation is equality at the base
type. We extend quote and unquote to produce witnesses and require a
witness of this logical relation, respectively. This is depicted in
figure \ref{fig:diagramold}. The commutativity of the right hand
triangle gives completeness: starting with a term, a semantic value
and a witness that these are related, we get a normal form, and then
if we embed it back into terms, we get a term equal to the one we
started with.

Stability can be proven by mutual induction on terms and normal forms.

\begin{figure}[!h]
\centering
\begin{tikzpicture}
\node (NE) at (0,0) {$\NE_A$};
\node (XX) at (4,0) {$\Sigma\,(\TM_A \times \ll A\rr)\,\R_A$};
\node (NF) at (8,0) {$\NF_A$};
\node (TM) at (4,-2) {$\TM_A$};
\draw[->] (NE) edge node[above] {$\u'_A$} (XX);
\draw[->] (XX) edge node[above] {$\q'_A$} (NF);
\draw[->] (NE) edge node[below] {$\cul\blank\cur\hspace{1em}$} (TM);
\draw[->] (NF) edge node[below] {$\hspace{1em}\cul\blank\cur$} (TM);
\draw[->] (XX) edge node[right] {$\mathsf{proj}$} (TM);
\end{tikzpicture}
\caption{The type of quote and unquote for a type $A$ in NBE for
  simple types. We use primed notations for the unquote and quote
  functions to denote that they include the completeness proof. This
  is a diagram in the category of presheaves.}
\label{fig:diagramold}
\end{figure}

A nice property of this normalisation proof is that the part of
unquote (and quote) which gives (and uses) $\ll A\rr$ can be defined
separately from the part which gives relatedness. This means that the
normalisation function can be defined independently from the proof
that it is complete.

\subsection{NBE for type theory}

In this subsection, we explain why the naive generalisation of the
proof for the simply typed case does not work in the presence of
dependent types and how we solve this problem.

In the case of simple type theory, types are closed, so they are
interpreted as presheaves just as contexts. When we have dependent
types, types depend on contexts, hence they are interpreted as
families of presheaves in the presheaf model (we omit functoriality).
\begin{alignat*}{3}
  & \ll\Gamma\rr && : |\REN| \ra \Set \\
  & \ll \Gamma \vdash A \rr && : (\Psi:|\REN|)\ra \ll\Gamma\rr\,\Psi \ra \Set
\end{alignat*}
We can declare quote for contexts the same way as for simple types,
but quote for types has to be more subtle. Our first candidate is the
following where it depends on quote for contexts (we omit the
naturality properties).
\begin{alignat*}{5}
  & \q_\Gamma && : (\Psi:|\REN|) \ra \ll\Gamma\rr\,\Psi \ra \Tms\,\Psi\,\Gamma \\
  & \q_{\Gamma\vdash A} && : (\Psi:|\REN|)(\alpha : \ll\Gamma\rr\,\Psi) \ra \ll A\rr_\Psi\,\alpha \ra \Nf\,\Psi\,\big(A[\q_{\Gamma, \Psi}\,\alpha]\big)
\end{alignat*}
The type of unquote also depends on quote for contexts.
\begin{alignat*}{5}
  & \u_{\Gamma\vdash A} && : (\Psi:|\REN|)(\alpha : \ll\Gamma\rr\,\Psi) \ra \Ne\,\Psi\,\big(A[\q_{\Gamma, \Psi}\,\alpha]\big) \ra \ll A\rr\,\Psi\,\alpha
\end{alignat*}
When we try to define quote and unquote following this specification,
we observe that we need some new equations to typecheck our
definition. E.g.\ quote for function types needs that quote after
unquote is the identity up to embedding:
$\cul\blank\cur\circ\q_A\,\circ\u_A \equiv \cul\blank\cur$. This is
however the consequence of the logical relation between the syntax and
the presheaf model: we can read it off figure \ref{fig:diagramold} by
the commutativity of the diagram: if we embed a neutral term into
terms, it is the same as unquoting, then quoting, then embedding.

Hence, our second attempt is defining quote and unquote mutually with
their correctness proofs. It is not very surprising that when moving
to dependent types the well-typedness of normalisation depends on
completeness. The types of quote and unquote become the following.
\begin{alignat*}{4}
  & \q_{\Gamma\vdash A} && : \, && (\Psi:|\REN|)(\rho:\Tms\,\Psi\,\Gamma)(\alpha:\ll\Gamma\rr\,\Psi)(p : \R_{\Gamma\,\Psi}\,\rho\,\alpha) \\
  & && && (t : \Tm\,\Psi\,A[\rho])(v : \ll A\rr_\Psi\,\alpha) \ra \R_{A\,\Psi\,p}\,t\,v \ra \Sigma(n:\NF_A\,\rho).t\equiv\cul n\cur \\
  & \u_{\Gamma\vdash A} && && \Psi\,\rho\,\alpha\,p : (n : \Ne\,\Psi\,A[\rho]) \ra \Sigma(v : \ll A\rr_\Psi\,\alpha).\R_{A\,\Psi\,p}\,\cul n\cur\,v
\end{alignat*}
However there seems to be no way to define quote and unquote this way
because quote does not preserve the logical relation. The problem is
that when defining unquote at $\Pi$ we need to define a semantic
function which works for arbitrary inputs, not only those which are
related to a term. The first component of unquote at $\Pi$ has the
following type.
\begin{alignat*}{4}
  & \proj_1\Big(\u_{\Gamma\vdash \Pi\,A\,B}\,\Psi\,\rho\,\alpha\,p\,\big(n : \Ne\,\Psi\,(\Pi\,A\,B)[\rho]\big)\Big) \\
  & \hspace{1em} : \forall\Omega.(\beta:\REN(\Omega,\Psi))\big(x:\ll A\rr_\Omega\,(\ll\Gamma\rr\,\beta\,\alpha)\big) \ra \ll B\rr_\Omega\,(\ll\Gamma\rr\,\beta\,\alpha, x)
\end{alignat*}
We should define this as unquoting the application of the neutral
function $n$ and quoting the input $x$. However we can't quote an
arbitrary semantic $x$, we also need a witness that it is related to a
term. It seems that we have to restrict the presheaf model to only
contain semantic elements which are related to some term.

Indeed, this is our solution: we merge the presheaf model and the
logical relation into a single proof-relevant logical predicate. We
denote the logical predicate at a context $\Gamma$ by $\r{\Gamma}$. We
define normalisation following the diagram in figure
\ref{fig:diagramnew}.

\begin{figure}[h]
\centering
\begin{tikzpicture}
\node (NE) at (0,0) {$\NE_\Gamma$};
\node (XX) at (4,0) {$\Sigma\,\TM_\Gamma\,\r{\Gamma}$};
\node (NF) at (8,0) {$\NF_\Gamma$};
\node (TM) at (4,-2) {$\TM_\Gamma$};
\draw[->] (NE) edge node[above] {$\u_\Gamma$} (XX);
\draw[->] (XX) edge node[above] {$\q_\Gamma$} (NF);
\draw[->] (NE) edge node[below] {$\cul\blank\cur\hspace{1em}$} (TM);
\draw[->] (NF) edge node[below] {$\hspace{1em}\cul\blank\cur$} (TM);
\draw[->] (XX) edge node[right] {$\mathsf{proj}$} (TM);
\end{tikzpicture}
\caption{The types of quote and unquote for a context $\Gamma$ in our
  proof.}
\label{fig:diagramnew}
\end{figure}

In the presheaf model, the interpretation of the base type was normal
forms at the base type and the logical relation at the base type was
equality of the term and the normal form. In our case, the logical
predicate at the base type will say that there exists a normal form
which is equal to the term (this is why it needs to be
proof-relevant). This solves the problem mentioned before: now the
semantics of a term will be the same term together with a witness of
the predicate for that term.

\subsection{Structure of the proof and the paper}

In this subsection, we give a high level sketch of the proof. Sections
\ref{sec:object_theory}, \ref{sec:inj}, \ref{sec:REN}, \ref{sec:Nf}
are fully formalised in Agda, the computational parts of sections
\ref{sec:logpred}, \ref{sec:quote} and \ref{sec:fruits} are
formalised, but some of the naturality and functoriality properties
are left as holes. The formalisation is available online
\cite{supplLMCS}. The proofs are available in full detail on paper
(including everything that we omitted in this paper and which is not
finished in the formalisation) in the second author's thesis
\cite{kaposiphd}.

In section \ref{sec:metatheory} we briefly summarize the metatheory we
are working in.

In section \ref{sec:object_theory} we define the syntax for type
theory as a quotient inductive inductive type (QIIT)
\cite{ttintt}. The arguments of the eliminator for the QIIT form a
model of type theory.

In section \ref{sec:inj} we prove injectivity of context extension and
the type formers $\El$ and $\Pi$. We will need these for proving
decidability of equality for normal forms.

In section \ref{sec:REN} we define the category of renamings $\REN$:
objects are contexts and morphisms are renamings.

In section \ref{sec:logpred} we define the proof-relevant presheaf
logical predicate interpretation of the syntax. The interpretation has
$\REN$ as the base category and two parameters for the interpretations
of $\U$ and $\El$. This interpretation can be seen as a dependent
version of the presheaf model of type theory. E.g.\ a context in the
presheaf model is interpreted as a presheaf. Now it is a family of
presheaves dependent on a substitution into that context. The
interpretations of base types can depend on the actual elements of the
base types. The interpretation of substitutions and terms are what are
usually called fundamental theorems.

In section \ref{sec:Nf} we define neutral terms and normal forms
together with their renamings and embeddings into the syntax ($\cul
\blank \cur$). With the help of these, we define the interpretations
of $\U$ and $\El$. The interpretation of $\U$ at a term of type $\U$
will be a neutral term of type $\U$ which is equal to the term. We
also prove decidability of equality for normal forms.

In section \ref{sec:quote} we mutually define the natural
transformations quote and unquote. We define them by induction on
contexts and types as shown in figure \ref{fig:diagramnew}. Quote
takes a term and a semantic value at that term into a normal term and
a proof that the normal term is equal to it. Unquote takes a neutral
term into a semantic value at the neutral term.

Finally, in section \ref{sec:fruits}, we put together the pieces by
defining the normalisation function and showing that it is complete
and stable. In addition, we show decidability of equality and
consistency.

\subsection{Related work}

Normalisation by evaluation was first formulated by Schwichtenberg and
Berger \cite{berger1991inverse}, subsequently a categorical account
using presheaf categories was given \cite{alti:ctcs95} and this
approach was extended to System F \cite{alti:lics96,alti:f97} and
coproducts \cite{alti:lics01}. The present work can be seen as a
continuation of this line of research. A fully detailed description of
our proof can be found in the PhD thesis of the second author
\cite{kaposiphd}.

The term normalisation by evaluation is also more generally used to
describe semantic based normalisation functions. E.g.\ Danvy is using
semantic normalisation for partial evaluation \cite{danvy1999type}.
Normalisation by evaluation using untyped realizers has been applied
to dependent types by Abel et al.
\cite{abel2007normalization,abel2010towards,abel2013normalization}.
Danielsson \cite{danielsson2006formalisation} has formalized NBE for
dependent types but he doesn't prove soundness of normalisation.

Our proof of injectivity of type formers is reminiscent in
\cite{harperpfenning} and the proof of decidability of normal forms is
similar to that of \cite{abelscherer}.


\section{Metatheory and notation}
\label{sec:metatheory}

We are working in intensional Martin-L{\"o}f Type Theory with
postulated extensionality principles using Agda as a vehicle
\cite{agda, agdawiki}. We make use of quotient inductive inductive
types (QIITs, see section 6 of \cite{ttintt}). QIITs are a
combiniation of inductive inductive types \cite{forsberg-phd} and
higher inductive types \cite{book}. The metatheory of QIITs is not
developed yet, however we hope that they can be justified by a setoid
model \cite{alti:lics99}. We only use one instance of a QIIT, the
definition of the syntax. We extend Agda with this QIIT using axioms
and rewrite rules \cite{cockxsprinkles}. The usage of rewrite rules
guarrantees that injectivity and disjointness of constructors of the
QIIT are not available to the unification mechanisms of Agda. Also,
pattern matching on constructors of the QIIT is not available, the
only way to define a function from the QIIT is to use the eliminator.

When defining an inductive type $A$, we first declare the type by
$\data\,A:S$ where $S$ is the sort, then we list the constructors. For
inductive inductive types we first declare all the types, then
following a second $\data$ keyword we list the constructors. We also
postulate functional extensionality which is a consequence of having
an interval QIIT anyway. We assume $\mathsf{K}$, that is, we work in a
strict type theory.

We follow Agda's convention of denoting the universe of types by
$\Set$, we write function types as $(x:A) \ra B$ or $\forall x.B$,
implicit arguments are written in curly braces $\{x:A\} \ra B$ and can
be omitted or given in curly braces or lower index. If some arguments
are omitted, we assume universal quantification, e.g.\ $(y : B\,x) \ra
C$ means $\forall\,x.(y:B\,x) \ra C$ if $x$ is not given in the
context. We write $\Sigma(x:A).B$ for $\Sigma$ types. We overload
names e.g.\ the action on objects and morphisms of a functor is
denoted by the same symbol.

The identity type (propositional equality) is denoted
$\blank\equiv\blank$ and its constructor is $\refl$. Transport of a
term $u : P\,a$ along an equality $p : a\equiv a'$ is denoted
$\tr{p}{u} : P\,a'$. We denote $(\tr{p}{u}) \equiv u'$ by $u \equiv^p
u'$. We write $\ap$ for congruence, that is $\ap\,f\,p : f\,a \equiv
f\,a'$ if $p : a \equiv a'$. We write $\blank\trans\blank$ for
transitivity and $\blank^{-1}$ for symmetry of equality. For
readability, we will omit writing transports in the informal
presentation most of the time, that is, our informal notation is that
of extensional type theory. This choice is justified by the
conservativity of extensional type theory over intensional type theory
with $\mathsf{K}$ and functional extensionality
\cite{hofmann95conservativity,Oury2005}. This allows writing
e.g.\ $f\,a$ where $f : A \ra B$ and $a : A'$ in case there is an
equality in scope which justifies $A \equiv A'$.

Sometimes we use Coq-style definitions: we write $\mathsf{d}\,(x:A) :
B := t$ for defining $\mathsf{d}$ of type $(x : A) \ra B$ by $\lambda
x.t$. We also use Agda-style pattern matching definitions. We use the
underscore $\_$ to denote arguments that we don't need e.g. the
constant function is written $\mathsf{const}\,x\,\_ := x$.


\section{Object theory}
\label{sec:object_theory}

The object theory is a basic type theory with dependent function
space, an uninterpreted base type $\U$ and an uninterpreted family
over this base type $\El$. We use intrinsic typing (that is, we only
define well typed terms, term formers and derivation rules are
indentified), and we present the theory as a QIIT, that is, we add
conversion rules as equality constructors. We define an explicit
substitution calculus, hence substitutions are part of the syntax and
the syntax is purely inductive (as opposed to inductive
recursive). For a more detailed presentation, see \cite{kaposiphd}.

The syntax constitutes of contexts, types, substitutions and terms. We
declare the QIIT of the syntax as follows.
\begin{alignat*}{3}
  & \data \, \Con && : \Set \\
  & \data \, \Ty && : \Con \ra \Set \\
  & \data \, \Tms && : \Con \ra \Con \ra \Set \\
  & \data \, \Tm && : (\Gamma : \Con) \ra \Ty \, \Gamma \ra \Set
\end{alignat*}
We use the convention of naming contexts $\Gamma, \Delta, \Theta$,
types $A, B$, terms $t, u$, substitutions $\sigma, \nu, \delta$.

The point constructors are listed in the left column and the equality
constructors in the right.
\begin{alignat*}{6}
  & \hspace{-0.5em} \data && &&                                                                                                           \data \\										       
  & \hspace{-0.5em} \ind \cdot && : \Con && 												  \ind [\id] && : A[\id] \equiv A \\							       
  & \hspace{-0.5em} \ind \blank,\blank && : (\Gamma : \Con) \ra \Ty \, \Gamma \ra \Con && 						  \ind [][] && : A[\sigma][\nu] \equiv A[\sigma \circ \nu] \\				       
  & \hspace{-0.5em} \ind \blank[\blank] && : \Ty \, \Delta \ra \Tms \, \Gamma \, \Delta \ra \Ty \, \Gamma && 				  \ind \U[] && : \U[\sigma] \equiv \U \\							       
  & \hspace{-0.5em} \ind \U && : \Ty \, \Gamma && 											  \ind \El[] && : (\El\,\hat{A})[\sigma] \equiv \El\,(\tr{\U[]}{\hat{A}[\sigma]}) \\	       
  & \hspace{-0.5em} \ind \El && : \Tm \, \Gamma \, \U \ra \Ty \, \Gamma && 								  \ind \Pi[] && : (\Pi\,A\,B)[\sigma] \equiv \Pi\,(A[\sigma])\,(B[\sigma\uparrow A]) \\		       
  & \hspace{-0.5em} \ind \Pi && : (A : \Ty \, \Gamma) \ra \Ty \, (\Gamma , A) \ra \Ty \, \Gamma && 					  \ind \id\circ && : \id \circ \sigma \equiv \sigma \\					       
  & \hspace{-0.5em} \ind \id && : \Tms \, \Gamma \, \Gamma && 										  \ind \circid && : \sigma \circ \id \equiv \sigma \\					       
  & \hspace{-0.5em} \ind \blank\circ\blank && : \Tms\,\Theta\,\Delta \ra \Tms\,\Gamma\,\Theta \ra \Tms\,\Gamma\,\Delta && 		  \ind \circcirc && : (\sigma \circ \nu) \circ \delta \equiv \sigma \circ (\nu \circ \delta) \\  
  & \hspace{-0.5em} \ind \epsilon && : \Tms \, \Gamma \, \cdot && 									  \ind \epsilon\eta && : \{\sigma : \Tms\,\Gamma\,\cdot\} \ra \sigma \equiv \epsilon \\	       
  & \hspace{-0.5em} \ind \blank,\blank && : (\sigma : \Tms\,\Gamma\,\Delta) \ra \Tm\,\Gamma\,A[\sigma] \ra \Tms\,\Gamma\,(\Delta,A) && 	  \ind \pi_1\beta && : \pi_1\,(\sigma, t) \equiv \sigma \\				       
  & \hspace{-0.5em} \ind \pi_1 && : \Tms\,\Gamma\,(\Delta,A) \ra \Tms\,\Gamma\,\Delta && 						  \ind \pi\eta && : (\pi_1\,\sigma, \pi_2\,\sigma) \equiv \sigma \\			       
  & \hspace{-0.5em} \ind \blank[\blank] && : \Tm\,\Delta\,A \ra (\sigma : \Tms\,\Gamma\,\Delta) \ra \Tm\,\Gamma\,A[\sigma] && 		  \ind ,\circ && : (\sigma, t) \circ \nu \equiv (\sigma \circ \nu) , (\tr{[][]}{t[\nu]}) \\      
  & \hspace{-0.5em} \ind \pi_2 && : (\sigma : \Tms\,\Gamma\,(\Delta,A)) \ra \Tm\,\Gamma\,A[\pi_1 \, \sigma] && 				  \ind \pi_2\beta && : \pi_2\,(\sigma, t) \equiv^{\pi_1\beta} t \\			       
  & \hspace{-0.5em} \ind \lam && : \Tm\,(\Gamma,A)\,B \ra \Tm\,\Gamma\,(\Pi\,A\,B) && 							  \ind \Pi\beta && : \app\,(\lam\,t) \equiv t \\						       
  & \hspace{-0.5em} \ind \app && : \Tm\,\Gamma\,(\Pi\,A\,B) \ra \Tm\,(\Gamma,A)\,B &&						          \ind \Pi\eta && : \lam\,(\app\,t) \equiv t \\						       
  & && && 														  \ind \lam[] && : (\lam\,t)[\sigma] \equiv^{\Pi[]} \lam\,(t[\sigma\uparrow A])                          
\end{alignat*}
The constructors can be summarized as follows.
\begin{itemize}
\item Substitutions form a category with a terminal object. This
  includes the categorical substitution laws for types $[\id]$ and
  $[][]$. 
\item Substitution laws for types $\U[]$, $\El[]$, $\Pi[]$.
\item The laws of comprehension which state that we have the natural
  isomorphism
  \begin{equation*}
    \pi_1\beta, \pi_2\beta\,\,\arcfromtop \hspace{2em} \blank,\blank\downarrow\begin{array}{l}\infer={\Tms\,\Gamma\,(\Delta,A)}{\sigma : \Tms\,\Gamma\,\Delta && \Tm\,\Gamma\,A[\sigma]}\end{array}\uparrow \pi_1, \pi_2 \hspace{2em} \arcfrombottom\,\,\pi\eta
  \end{equation*}
  where naturality\footnote{If one direction of an isomorphism is
    natural, so is the other. This is why it is enough to state
    naturality for $\blank,\blank$ and not for $\pi_1$, $\pi_2$.} is
  given by $,\circ$.
\item The laws for function space which are given by the natural
  isomorphism
  \begin{equation*}
    \Pi\beta\,\,\arcfromtop \hspace{2em} \lam\downarrow\begin{array}{l}\infer={\Tm\,\Gamma\,(\Pi\,A\,B)}{\Tm\,(\Gamma,A)\,B}\end{array}\uparrow \app \hspace{2em} \arcfrombottom\,\,\Pi\eta
  \end{equation*}
  where naturality is given by $\lam[]$.
\end{itemize}

Note that the equality $\pi_2\beta$ lives over $\pi_1\beta$. Also, we
had to use transport to typecheck $\El[]$ and $,\circ$.
We used lifting of a substitution in the types of $\Pi[]$ and
$\lam[]$. It is defined as follows.
\begin{alignat*}{3}
  & \blank\uparrow\blank : (\sigma:\Tms\,\Gamma\,\Delta) \ra \Ty\,\Delta \ra \Tms\,(\Gamma,A[\sigma])\,(\Delta,A) \\
  & \sigma \uparrow A := (\sigma \circ \pi_1\,\id), (\tr{[][]}{\pi_2\,\id})
\end{alignat*}
We use the categorical $\app$ operator but the usual one
($\blank\$\blank$) can also be derived.
\begin{alignat*}{3}
  & \lb (u : \Tm\,\Gamma\,A) \rb && : \Tms\,\Gamma\,(\Gamma,A) && := \id, \tr{[\id]^{-1}}{u} \\
  & (t : \Tm\,\Gamma\,(\Pi\,A\,B)) \$ (u : \Tm\,\Gamma\,A) && : B[\lb u \rb] && := (\app\,t)[ \lb u \rb ]
\end{alignat*}

When we define a function from the above syntax, we need to use the
eliminator. The eliminator has four motives corresponding to what
$\Con$, $\Ty$, $\Tms$ and $\Tm$ get mapped to and one method for each
constructor including the equality constructors. The methods for point
constructors are the elements of the motives to which the constructor
is mapped. The methods for the equality constructors demonstrate
soundness, that is, the semantic constructions respect the syntactic
equalities. The eliminator comes in two different flavours: the
non-dependent and dependent version. In our constructions we use the
dependent version. The motives and methods for the non-dependent
eliminator (recursor) collected together form a model of type theory,
they are equivalent to Dybjer's Categories with Families
\cite{dybjer1996internal}.

To give an idea of what the eliminator looks like we list its motives
and some of its methods. For a complete presentation and an algorithm
for deriving these from the constructors, see \cite{kaposiphd}. As
names we use the names of the constructors followed by an upper index
$^\M$.
\begin{alignat*}{3}
  & \Con^\M && : \Con \ra \Set \\
  & \Ty^\M && : (\Con^\M\,\Gamma) \ra \Ty\,\Gamma \ra \Set \\
  & \Tms^\M && : (\Con^\M\,\Gamma) \ra (\Con^\M\,\Delta) \ra \Tms\,\Gamma\,\Delta \ra \Set \\
  & \Tm^\M && : (\Gamma^\M : \Con^\M\,\Gamma) \ra \Ty^\M\,\Gamma^\M\,A \ra \Tm\,\Gamma\,A \ra \Set \\
  & \cdot^\M && : \Con^\M\,\cdot \\
  & \blank,^\M\blank && : (\Gamma^\M : \Con^\M\,\Gamma) \ra \Ty^\M\,\Gamma^\M\,A \ra \Con^\M\,(\Gamma,A) \\
  & \id^\M && : \Tms^\M\,\Gamma^\M\,\Gamma^\M\,\id \\
  & \blank\circ^\M\blank && : \Tms^\M\,\Theta^\M\,\Delta^\M\,\sigma \ra \Tms^\M\,\Gamma^\M\,\Theta^\M\,\nu \ra \Tms^\M\,\Gamma^\M\,\Delta^\M\,(\sigma\circ\nu) \\
  & \circid^\M && : \sigma^\M \circ^\M \id^\M \equiv^{\circ\id} \sigma^\M \\
  & \pi_2\beta^\M && : \pi_2^\M\,(\rho^\M{,}^\M t^\M) \equiv^{\pi_1\beta^\M, \pi_2\beta} t^\M
\end{alignat*}
Note that the method equality $\circ\id^\M$ lives over the constructor
$\circ\id$ while the method equality $\pi_2\beta^\M$ lives both over
the method equality $\pi_1\beta^\M$ and the equality constructor
$\pi_2\beta$.

There are four eliminators for the four constituent types. These are
understood in the presence of all the motives and methods given above.
\begin{alignat*}{5}
  & \Elim\Con && : (\Gamma : \Con) && \ra \Con^\M\,\Gamma \\
  & \Elim\Ty && : (A : \Ty\,\Gamma) && \ra \Ty^\M\,(\Elim\Con\,\Gamma)\,A \\
  & \Elim\Tms && : (\sigma : \Tms\,\Gamma\,\Delta) && \ra \Tms^\M\,(\Elim\Con\,\Gamma)\,(\Elim\Con\,\Delta)\,\sigma \\
  & \Elim\Tm && : (t : \Tm\,\Gamma\,A) && \ra \Tm^\M\,(\Elim\Con\,\Gamma)\,(\Elim\Ty\,A)\,t
\end{alignat*}
We have the usual $\beta$ computation rules such as the following.
\begin{alignat*}{5}
 \Elim\Con\,(\Gamma,A) & = \Elim\Con\,\Gamma ,^\M \Elim\Ty\,A \\
 \Elim\Tms\,(\sigma\circ\nu) & = \Elim\Tms\,\sigma \circ^\M \Elim\Tms\,\nu
\end{alignat*}
There are no $\beta$ rules for the equality constructors (such rules
would be only interesting in a setting without $\mathsf{K}$).


\section{Injectivity of context and type formers}
\label{sec:inj}

As examples of using the eliminator we prove injectivity of context
and type constructors. We will need these results when proving
decidability of equality for normal forms in section \ref{sec:Nf}.

We start by injectivity of context extension. As there are no equality
constructors for contexts, the proof follows the usual argument for
injectivity of constructors for inductive types.

First, given $\Gamma_0 : \Con$ and $A_0 : \Ty\,\Gamma_0$ we define a
family over contexts $\mathsf{P} : \Con \ra \Set$ using the
eliminator. We specify the motives and methods as follows.
\begin{alignat*}{5}
  & \Con^\M\,\_ && := \Set \\
  & \Ty^\M\,\_\,\_ && := \top \\
  & \Tms^\M\,\_\,\_\,\_ && := \top \\
  & \Tm^\M\,\_\,\_\,\_ && := \top \\
  & \cdot^\M && := \bot \\
  & \blank,^\M\blank\,\{\Gamma : \Con\}\,\_\,\{A : \Ty\,\Gamma\}\,\_ && := \Sigma(q:\Gamma_0\equiv \Gamma).A_0 \equiv^q A \\
  & \id^\M, ... && := \tt \\
  & \circid^\M, ... && := \refl
\end{alignat*}
A context is interpreted as a type. Types, substitutions and terms are
interpreted as elements of the unit type, hence the interpretations of
all the type formers, substitution and term constructors are trivially
$\tt$ and all the equalities hold by reflexivity. The empty context is
interpreted as the empty type (we will never need this later) and an
extended context $(\Gamma,A)$ is interpreted as a pair of equalities
between $\Gamma_0$ and $\Gamma$ and $A_0$ and $A$ (the latter depends
on the former equality). When defining $\blank,^\M\blank$ we wrote
underscores for the interpretations of $\Gamma$ and $A$ (having types
$\Set$ and $\top$, respectively) thus ignoring these arguments, we
only used $\Gamma$ and $A$ themselves which are implicit arguments of
the eliminator. Using the above motives and methods, we define $\P :=
\Elim\Con : \Con\ra\Set$ and the $\beta$ rule tells us that
\[
\P\,(\Gamma_0,A_0) = \Sigma(q:\Gamma_0\equiv \Gamma_0).A_0 \equiv^q A_0
\]
and 
\[
\P\,(\Gamma_1,A_1) = \Sigma(q:\Gamma_0\equiv \Gamma_1).A_0 \equiv^q A_1.
\]
We can prove the first one by $(\refl, \refl)$ and given an equality
$w$ between the indices $(\Gamma_0,A_0)$ and $(\Gamma_1,A_1)$, we can
transport it to the second one. This proves injectivity:
\[
\inj, : \big(w: (\Gamma_0,A_0) \equiv (\Gamma_1,A_1)\big) := \tr{w}{(\refl,\refl)} : \Sigma(q:\Gamma_0\equiv \Gamma_1).A_0 \equiv^q A_1
\]

To show injectivity of type formers, we start by the definition of
normal types. These are either $\U$, $\El$ or $\Pi$, but not
substituted types. Then we show normalisation of types using the
eliminator (this just means pushing down the substitutions until we
reach a $\U$ or $\El$). Finally we prove the first injectivity lemma
for $\Pi$ using normalisation.

Normal types are given by the following indexed inductive type which
is defined mutually with the embedding back into types. In spite of
their name, these are not fully normal types: they can include
arbitrary non-normal terms through $\El$. Note that we use overloaded
constructor names.
\begin{alignat*}{10}
  & \data \, \NTy && : (\Gamma:\Con) \ra \Set \\
  & \cul \blank \cur && : \NTy\,\Gamma \ra \Ty\,\Gamma \\
  & \data \, \NTy && \\
  & \ind \U && : \NTy\,\Gamma \\
  & \ind \El && : \Tm\,\Gamma\,\U \ra \NTy\,\Gamma \\
  & \ind \Pi && : (A : \NTy\,\Gamma)\ra \NTy\,(\Gamma,\cul A\cur)\ra \NTy\,\Gamma \\
  & \cul \Pi\,A\,B \cur && := \Pi\,\cul A\cur\,\cul B\cur \\
  & \cul \U \cur && := \U \\
  & \cul \El\,\hat{A} \cur && := \El\,\hat{A}
\end{alignat*}
Substitution of normal types can be defined by ignoring the
substitution for $\U$, applying it to the term for $\El$ and
substituting recursively for $\Pi$. We need to mutually prove a lemma
saying that the embedding is compatible with substitution. As $\NTy$
is a simple inductive type (no equality constructors) we use pattern
matching notation when defining these functions.
\begin{alignat*}{10}
  & \blank[\blank] && : \NTy\,\Delta \ra \Tms\,\Gamma\,\Delta\ra \NTy\,\Gamma \\
  & \cul[]\cur && : (A : \NTy\,\Delta)(\sigma : \Tms\,\Gamma\,\Delta) \ra \cul A\cur[\sigma] \equiv \cul A[\sigma] \cur \\
  & (\Pi\,A\,B)[\sigma] && := \Pi\,(A[\sigma])\,(B[\tr{(\cul[]\cur\,A\,\sigma)}{\sigma\uparrow A}]) \\
  & \U[\sigma] && := \U \\
  & (\El\,\hat{A})[\sigma] && := \El\,(\hat{A}[\sigma]) \\
  & \cul[]\cur\,(\Pi\,A\,B)\,\sigma && := \Pi[] \trans \ap\Pi\,\big(\cul[]\cur\,A\,\sigma\big)\,\big(\cul[]\cur\,B\,(\sigma\uparrow A)\big) \\
  & \cul[]\cur\,\U\,\sigma && := {\U[]} \\
  & \cul[]\cur\,(\El\,\hat{A})\,\sigma && := {\El[]}
\end{alignat*}
When defining substitution of $\Pi$, we need to use $\cul[]\cur$ to
transport the lifted substitution $\sigma\uparrow A$ to the expected
type. The lemma $\cul[]\cur$ is proved using the substitution laws of
the syntax and the induction hypothesis in the case of $\Pi$. $\ap\Pi$
denotes the congruence rule for $\Pi$, its type is $(p_A : A\equiv A')
\ra B \equiv^{p_A} B' \ra \Pi\,A\,B\equiv \Pi\,A'\,B'$.

By induction on normal types, we prove the following two lemmas as
well.
\begin{alignat*}{10}
  & [\id] && : (A : \NTy\,\Gamma) \ra A[\id] \equiv A \\
  & [][] && : (A : \NTy\,\Gamma).\forall \sigma\,\nu. A[\sigma][\nu] \equiv A[\sigma\circ\nu]
\end{alignat*}

Now we can define the model of normal types using the following
motives for the eliminator.
\begin{alignat*}{10}
  & \Con^\M\,\_ && := \top \\
  & \Ty^\M\,\{\Gamma\}\,\_\,A && := \Sigma(A':\NTy\,\Gamma).A\equiv \cul A'\cur \\
  & \Tms^\M\,\_\,\_\,\_ && := \top \\
  & \Tm^\M\,\_\,\_\,\_ && := \top
\end{alignat*}
That is, the eliminator will map a type to a normal type and a proof
that the embedding of the normal type is equal to the original
type. Contexts, substitutions and terms are mapped to the trivial
type. Hence, the methods for contexts, substitutions and terms will be
all trivial and the equality methods for them can be proven by
$\refl$.

The methods for types are given as follows.
\begin{alignat*}{10}
  & \blank[\blank]^\M\,\{A\}\,(A', p_A)\,\{\sigma\}\,\_ && := \big(A'[\sigma], \ap\,(\blank[\sigma])\,p_A\trans\cul[]\cur\,A'\,\sigma\big) \\
  & \U^\M && := (\U, \refl) \\
  & \El^\M\,\{\hat{A}\}\,\_ && := (\El\,\hat{A}, \refl) \\
  & \Pi^\M\,\{A\}\,(A', p_A)\,\{B\}\,(B', p_B) && := \big(\Pi\,A'\,(\tr{p_A}\,B'), \ap\Pi\,p_A\,p_B\big)
\end{alignat*}
$\blank[\blank]^\M$ receives a type $A$ as an implicit argument, a
normal type $A'$ and a proof $p_A$ that they are equal, a substitution
$\sigma$ as an implicit argument and the semantic version of the
substitution which does not carry information. We use the above
defined $\blank[\blank]$ for substituting $A'$ and we need the
concatenation of the equalities $p_A$ and $\cul[]\cur$ to provide the
equality $\cul A'[\sigma]\cur \equiv \cul A[\sigma]\cur$. Mapping $\U$
and $\El\,\hat{A}$ to normal types is trivial, while in the case of
$\Pi\,A\,B$ we use the inductive hypotheses $A'$ and $B'$ to construct
$\Pi\,A'\,B'$, and in a similar way we use $p_A$ and $p_B$ to
construct the equality.

When proving the equality methods $[\id]^\M, [][]^\M, \U[]^\M,
\El[]^\M$ and $\Pi[]^\M$, it is enough to show that the first
components of the pairs (the normal types) are equal, the $p_A$ proofs
will be equal by $\mathsf{K}$. The equality methods $[\id]^\M$ and
$[][]^\M$ are given by the above lemmas $[\id]$ and $[][]$. The
semantic counterparts of the substitution laws $\U[]$ and $\El[]$ are
trivial, while $\Pi[]^\M$ is given by a straightforward induction.

Using the eliminator, we define normalisation of types as follows.
\begin{alignat*}{10}
  & \norm\,(A : \Ty\,\Gamma) : \NTy\,\Gamma := \proj_1\,(\Elim\Ty\,A)
\end{alignat*}
We can also show completeness and stability of normalisation (see
section \ref{sec:spec} for this nomenclature).
\begin{alignat*}{10}
  & \mathsf{compl}\,(A : \Ty\,\Gamma) : A\equiv\cul \norm\,A\cur := \proj_2\,(\Elim\Ty\,A) \\
  & \mathsf{stab}\,(A' : \NTy\,\Gamma) : A'\equiv \norm\,\cul A\cur
\end{alignat*}
Stability is proven by a straightforward induction on normal types.

Injectivity of $\Pi^\NTy$ (the $\Pi$ constructor for normal types) is
proven the same way as we did for context extension: the family
$\mathsf{P}$ can be simply given by pattern matching as $\NTy$ doesn't
have equality constructors. Given a type $A_0$, we define $\mathsf{P}$
as follows.
\begin{alignat*}{10}
  & \mathsf{P} : \NTy\,\Gamma \ra \Set \\
  & \mathsf{P}\,(\Pi^\NTy\,A\,B) := A_0 \equiv A \\
  & \mathsf{P}\,X := \bot
\end{alignat*}
Note that we can't define the same family over $\Ty$ (using the
eliminator of the syntax) because it does not respect the equality
$\Pi[]$. With the help of this $\mathsf{P}$, we can prove injectivity
by transporting the reflexivity proof of
$\mathsf{P}\,(\Pi^\NTy\,A_0\,B_0) = (A_0 \equiv A_0)$ to that of
$\mathsf{P}\,(\Pi^\NTy\,A_1\,B_1) = (A_0 \equiv A_1)$.
\[
  \mathsf{inj}\Pi^\NTy (w : \Pi^\NTy\,A_0\,B_0 \equiv \Pi^\NTy\,A_1\,B_1) : A_0\equiv A_1 := \tr{w}{\refl}
\]

We put together these pieces to prove injectivity of $\Pi^\Ty$ with
the diagram in figure \ref{fig:diagram_injpi}. We start with a proof
$p : \Pi^\Ty\,A_0\,B_0 \equiv \Pi^\Ty\,A_1\,B_1$, then use
completeness to get a proof $q : \cul\norm\,(\Pi^\Ty\,A_0\,B_0)\cur
\equiv \cul\norm\,(\Pi^\Ty\,A_1\,B_1)\cur$. Applying $\norm$ to both
sides and using stability we get $r : \norm\,(\Pi^\Ty\,A_0\,B_0)
\equiv \norm\,(\Pi^\Ty\,A_1\,B_1)$. The type of $r$ reduces to
$\Pi^\NTy\,(\norm\,A_0)\,(\norm\,B_0) \equiv
\Pi^\NTy\,(\norm\,A_1)\,(\norm\,B_1)$ and now we can apply the
injectivity of normal $\Pi$ to get that $\norm\,A_0 \equiv
\norm\,A_1$. As a last step we apply $\cul\blank\cur$ to both sides of
this equality and use completeness on $A_0$ and $A_1$ to obtain $A_0
\equiv A_1$.

\begin{figure}[h]
\centering
\begin{tikzpicture}
\node (x00) at (0,0) {$\Pi^\Ty\,A_0\,B_0$};
\node (x10) at (6,0) {$\Pi^\Ty\,A_1\,B_1$};
\draw[-] (x00) edge node[above] {$p$} (x10);
\node (x01) at (0,-2) {$\cul\norm\,(\Pi^\Ty\,A_0\,B_0)\cur$};
\node (x11) at (6,-2) {$\cul\norm\,(\Pi^\Ty\,A_1\,B_1)\cur$};
\draw[-] (x00) edge node[left] {$\mathsf{compl}\,(\Pi\,A_0\,B_0)$} (x01);
\draw[-] (x10) edge node[right] {$\mathsf{compl}\,(\Pi\,A_0\,B_0)$} (x11);
\draw[-,dashed] (x01) edge node[above] {$q$} (x11);
\node (x02) at (0,-3.5) {$\norm\,\cul\norm\,(\Pi^\Ty\,A_0\,B_0)\cur$};
\node (x12) at (6,-3.5) {$\norm\,\cul\norm\,(\Pi^\Ty\,A_1\,B_1)\cur$};
\draw[-] (x02) edge node[above] {$\ap\,\norm\,q$} (x12);
\node (x03) at (0,-5.5) {$\norm\,(\Pi^\Ty\,A_0\,B_0)$};
\node (x13) at (6,-5.5) {$\norm\,(\Pi^\Ty\,A_1\,B_1)$};
\draw[-] (x02) edge node[left] {$\mathsf{stab}\,\big(\norm\,(\Pi^\Ty\,A_0\,B_0)\big)$} (x03);
\draw[-] (x12) edge node[right] {$\mathsf{stab}\,\big(\norm\,(\Pi^\Ty\,A_1\,B_1)\big)$} (x13);
\node (x04) at (0,-7) {$\Pi^\NTy\,(\norm\,A_0)\,(\norm\,B_0)$};
\node (x14) at (6,-7) {$\Pi^\NTy\,(\norm\,A_1)\,(\norm\,B_1)$};
\draw[-] (x03.263) edge node {$$} (x04.97);
\draw[-] (x03.277) edge node {$$} (x04.83);
\draw[-] (x13.263) edge node {$$} (x14.97);
\draw[-] (x13.277) edge node {$$} (x14.83);
\draw[-,dashed] (x04) edge node[above] {$r$} (x14);
\node (x05) at (0,-8.5) {$\norm\,A_0$};
\node (x15) at (6,-8.5) {$\norm\,A_1$};
\draw[-] (x05) edge node[above] {${\mathsf{inj}\Pi^\NTy}\,r$} (x15);
\node (x06) at (0,-10) {$\cul\norm\,A_0\cur$};
\node (x16) at (6,-10) {$\cul\norm\,A_1\cur$};
\draw[-] (x06) edge node[above] {$\ap\,\cul\blank\cur\,\big({\mathsf{inj}\Pi^\NTy}\,r\big)$} (x16);
\node (x07) at (0,-12) {$A_0$};
\node (x17) at (6,-12) {$A_1$};
\draw[-] (x06) edge node[left] {$\mathsf{compl}\,A_0$} (x07);
\draw[-] (x16) edge node[right] {$\mathsf{compl}\,A_1$} (x17);
\draw[-,dashed] (x07) edge node[above] {${\mathsf{inj}\Pi^\Ty}\,p$} (x17);
\end{tikzpicture}
\caption{Proof of injectivity of $\Pi^\Ty$ in the domain. The dashed
  lines are given by the fillers of the squares. The double lines are
  definitional equalities.}
\label{fig:diagram_injpi}
\end{figure}

We only state the other two injectivity lemmas, they can be proved
analogously to $\inj\Pi^\Ty$.
\begin{alignat*}{5}
  & {\inj\Pi^\Ty}' && : (w : \Pi^\Ty\,A_0\,B_0 \equiv \Pi^\Ty\,A_1\,B_1) \ra B_0 \equiv^{\inj\Pi^\Ty\,w} B_1 \\
  & \inj\El && : \El^\Ty\,\hat{A_0} \equiv \El^\Ty\,\hat{A_1} \ra \hat{A_0} \equiv \hat{A_1}
\end{alignat*}

The above proof works for our small type theory but it is not obvious
how it would scale to a type theory with large elimination. In that
case the injectivity proof would depend on normalisation of terms.


\section{The category of renamings}
\label{sec:REN}

In this section we define the category of renamings $\REN$. Objects in
this category are contexts, morphisms are renamings of variables
($\Vars$).

We define typed De Bruijn variables $\Var$ and renamings $\Vars$
together with their embeddings into substitutions.
\begin{alignat*}{3}
  & \data \, \Var && : (\Psi : \Con) \ra \Ty\,\Psi \ra \Set \\
  & \ind \vze && : \Var\,(\Psi,A)\,(A[\pi_1\,\id]) \\
  & \ind \vsu && : \Var\,\Psi\,A \ra \Var\,(\Psi,B)\,(A[\pi_1\,\id]) \\
  & \cul \blank \cur && : \Vars\,\Omega\,\Psi \ra \Tms\,\Omega\,\Psi \\
  & \data \, \Vars && : \Con \ra \Con \ra \Set \\
  & \ind \epsilon && : \Vars\,\Psi\,\cdot \\
  & \ind \blank,\blank && : (\beta : \Vars\,\Omega\,\Psi) \ra \Var\,\Omega\,A[\cul \beta \cur] \ra \Vars\,\Omega\,(\Psi,A) \\
  & \cul \blank \cur && : \Var\,\Psi\,A \ra \Tm\,\Psi\,A \\
  & \cul \vze \cur && := \pi_2\,\id \\
  & \cul \vsu\,x\cur && := \cul x \cur [ \pi_1\,\id ]\\
  & \cul\epsilon\cur && := \epsilon \\
  & \cul\beta,x\cur && := \cul \beta \cur, \cul x \cur
\end{alignat*}
Variables are typed de Bruijn indices. $\vze$ projects out the last
element of the context, $\vsu$ extends the context, and the type $A :
\Ty\,\Psi$ needs to be weakened in both cases because we need to
interpret it in $\Psi$ extended by another type. Renamings are lists
of variables with the appropriate types. Embedding of variables into
terms uses the projections and the identity substitution, and
embedding renamings is pointwise.

We use the names $\Psi, \Omega, \Xi$ for objects of $\REN$, $x, y$ for
variables, $\beta, \gamma$ for renamings.

We need identity and composition of renamings for the categorical
structure. To define them, we also need weakening and renaming of
variables together with laws relating their embeddings to terms. We
only list the types as the definitions are straightforward
inductions. 
\begin{alignat*}{5}
  & \wkV && : \Vars\,\Omega\,\Psi \ra \Vars\,(\Omega,A)\,\Psi && \cul\wkV\cur && : \cul\beta\cur\circ\pi_1\id \equiv \cul\wkV\,\beta\cur \\
  & \id && : \Vars\,\Psi\,\Psi && \cul\id\cur && : \cul\id\cur \equiv \id \\
  & \blank\circ\blank && : \Vars\,\Xi\,\Psi \ra \Vars\,\Omega\,\Xi \ra \Vars\,\Omega\,\Psi && \cul\circ\cur && : \cul\beta\cur\circ\cul\gamma\cur \equiv \cul\beta\circ\gamma\cur \\
  & \blank[\blank] && : \Var\,\Psi\,A \ra (\beta : \Vars\,\Omega\,\Psi) \ra \Var\,\Omega\,A[\cul\beta\cur] \hspace{3em} && \cul[]\cur && : \cul x \cur[\cul\beta\cur] \equiv \cul x[\beta]\cur
\end{alignat*}
Renamings form a category, we omit the statement and proofs of the
categorical laws.


\section{The logical predicate interpretation}
\label{sec:logpred}

In this section, we define the proof-relevant presheaf logical
predicate interpretation of the type theory given in section
\ref{sec:object_theory}. It can be seen as a dependent version of the
presheaf model of type theory \cite{hofmann1997syntax}: for example,
in the presheaf model contexts are interpreted as presheaves (a
contravariant functor into $\Set$, that is, a set for each object with
some more structure). Now these sets depend not only on the object,
but also on a substitution into the corresponding context. The
categorically inclined reader can view this construction as the
categorical glueing \cite{crole} over the Yoneda embedding from the
term model to the presheaf model.

We start by defining the Yoneda embedding $\TM$ which embeds a context
into the sets of substitutions into that context, a type into the sets
of terms into that context, and substitutions and terms into maps
between the corresponding sets. We denote contravariant presheaves
over $\C$ by $\PSh\,\C$, families of presheaves over a presheaf by
$\FamPSh$ and natural transformations and sections by $\nat$ and $\S$,
respectively. See appendix \ref{appendix} for the definitions of these
categorical notions.
\begin{alignat*}{10}
  & \hspace{-2em} \Delta:\Con                                && \TM_\Delta && : \PSh\,\REN                         && \TM_\Delta\,\Psi && := \Tms\,\Psi\,\Delta                && \TM_\Delta\,\beta\,\rho && := \rho\circ\cul\beta\cur \\
  & \hspace{-2em} A : \Ty\,\Gamma                            && \TM_A     && : \FamPSh\,\TM_\Gamma                 && \TM_{A\,\Psi}\,\rho && := \Tm\,\Psi\,A[\rho] \hspace{1em} && \TM_A\,\beta\,t && := t[\cul\beta\cur] \\
  & \hspace{-2em} \sigma : \Tms\,\Gamma\,\Delta \hspace{1em} && \TM_\sigma && : \TM_\Gamma\nat\TM_\Delta             && \TM_{\sigma\,\Psi}\,\rho && := \sigma\circ\rho \\
  & \hspace{-2em} t : \Tm\,\Gamma\,A                         && \TM_t     && : \TM_\Gamma \S \TM_A \hspace{1em}    && \TM_{t\,\Psi}\,\rho && := t[\rho]
\end{alignat*}
$\TM_\Delta$ is a presheaf over $\REN$. The functor laws hold as
$\rho\circ\cul\id\cur \equiv \rho$ and
$(\rho\circ\cul\beta\cur)\circ\cul\gamma\cur \equiv
\rho\circ\cul\beta\circ\gamma\cur$. $\TM_A$ is a family of presheaves
over $\TM_\Gamma$, and similarly we have $t[\cul\id\cur] \equiv t$ and
$t[\cul\beta\cur][\cul\gamma\cur] \equiv
t[\cul\beta\circ\gamma\cur]$. $\TM_\sigma$ is a natural
transformation, naturality is given by associativity:
$(\sigma\circ\rho)\circ\cul\beta\cur \equiv
\sigma\circ(\rho\circ\cul\beta\cur)$. $\TM_t$ is a section, it is
naturality law can be verified as $t[\rho][\cul\beta\cur] \equiv
t[\rho\circ\cul\beta\cur]$. $\TM$ can be seen as a (weak) morphism in
the category of models of type theory from the syntax into the
presheaf model.

The motives for the presheaf logical predicate interpretation are
given as families over the Yoneda embedding $\TM$. In contrast with
section \ref{sec:inj}, here we use a recursive notation for defining
the motives and methods.\footnote{For reference, the ``arguments for the eliminator'' notation for the
motives looks as follows.
\begin{alignat*}{10}
  & \Con^\M\,\Delta && := \FamPSh\,\TM_\Delta \\
  & \Ty^\M\,\Gamma^\M\,A && := \FamPSh\,\big(\Sigma\,(\Sigma\,\TM_\Gamma\,\TM_A)\,\Gamma^\M[\wk]\big) \\
  & \Tms^\M\,\Gamma^\M\,\Delta^\M\,\sigma && := \Sigma\,\TM_\Gamma\,\Gamma^\M \S \Delta^\M[\TM_\sigma][\wk] \\
  & \Tm^\M\,\Gamma^\M\,A^\M\,t && := \Sigma\,\TM_\Gamma\,\Gamma^\M \S A^\M[\TM_t\uparrow\Gamma^\M]
\end{alignat*}}
\begin{alignat*}{10}
  & \Delta && : \Con && \r{\Delta} && : \FamPSh\,\TM_\Delta \\
  & A && : \Ty\,\Gamma && \r{A} && : \FamPSh\,\big(\Sigma\,(\Sigma\,\TM_\Gamma\,\TM_A)\,\r{\Gamma}[\wk]\big) \\
  & \sigma && : \Tms\,\Gamma\,\Delta && \r{\sigma} && : \Sigma\,\TM_\Gamma\,\r{\Gamma} \S \r{\Delta}[\TM_\sigma][\wk] \\
  & t && : \Tm\,\Gamma\,A \hspace{2em} && \r{t} && : \Sigma\,\TM_\Gamma\,\r{\Gamma} \S \r{A}[\TM_t\uparrow\r{\Gamma}]
\end{alignat*}
Note that $\blank[\blank]$ above is the composition of a family of
presheaves with a natural transformation, $\wk$ is the weakening
natural transformation and $\blank\uparrow\blank$ is lifting of a
section (see appendix \ref{appendix} for details).

We unfold these definitions below following appendix \ref{appendix}.

A context $\Delta$ is mapped to a family of presheaves over
$\TM_\Delta$. That is, for every substitution $\rho :
\TM_\Delta\,\Psi$ we have a type $\r{\Delta}_\Psi\,\rho$ expressing
that the logical predicate holds for $\rho$. Moreover, we have the
renaming $\r{\Delta}\,\beta : \r{\Delta}_\Psi\,\rho \ra
\r{\Delta}_\Omega\,(\TM_\Delta\,\beta\,\rho)$ for a $\beta :
\REN(\Omega,\Psi)$. Sometimes we omit the parameter $_\Psi$, also for
the $\r{}$ operations on types, substitutions and terms.

$\r{A}$ is the logical predicate at a type $A$. It depends on a
substitution (for which the predicate needs to hold as well) and a
term. $\r{A}_\Psi\,(\rho, s, \alpha)$ expresses that the logical
predicate holds for term $s : \Tm\,\Psi\,A[\rho]$.
\begin{equation*}
  \infer{\r{A}_\Psi\,(\rho, s, \alpha) : \Set}{A : \Ty\,\Gamma && \Psi : |\REN| && \rho : \TM_\Gamma\,\Psi && s : \TM_A\,\rho && \alpha : \r{\Gamma}_\Psi\,\rho}
\end{equation*}
It is also stable under renamings. That is, for a $\beta : \REN(\Omega,\Psi)$ we have
\begin{equation*}
  \r{A}\,\beta : \r{A}_\Psi\,(\rho, s, \alpha) \ra \r{A}_\Omega\,(\TM_\Gamma\,\beta\,\rho, \TM_A\,\beta\,s, \r{\Gamma}\,\beta\,\alpha).
\end{equation*}

A substitution $\sigma$ is mapped to $\r{\sigma}$ which expresses the
fundamental theorem of the logical predicate at $\sigma$: for any
other substitution $\rho$ for which the predicate holds, we can
compose it with $\sigma$ and the predicate will hold for the
composition.
\begin{equation*}
  \infer{\r{\sigma}_\Psi\,(\rho, \alpha) : \r{\Delta}_\Psi\,(\sigma\circ\rho)}{\sigma : \Tms\,\Gamma\,\Delta && \Psi : |\REN| && \rho : \TM_\Gamma\,\Psi && \alpha : \r{\Gamma}_\Psi\,\rho}
\end{equation*}
The fundamental theorem is also natural.
\begin{equation*}
  \r{\Delta}\,\beta\,(\r{\sigma}_\Psi\,(\rho, \alpha)) \equiv \r{\sigma}_\Omega\,(\TM_\Gamma\,\beta\,\rho, \r{\Gamma}\,\beta\,\alpha)
\end{equation*}

A term $t$ is mapped to the fundamental theorem at the term: given a
substitution $\rho$ for which the predicate holds, it also holds for
$t[\rho]$ in a natural way.
\begin{equation*}
  \infer{\r{t}_\Psi\,(\rho, \alpha) : \r{A}_\Psi\,(\rho, t[\rho], \alpha)}{t : \Tm\,\Gamma\,A && \Psi : |\REN^\op| && \rho : \TM_\Gamma\,\Psi && \alpha : \r{\Gamma}_\Psi\,\rho}
\end{equation*}
\begin{equation*}
  \r{A}\,\beta\,\big(\r{t}_\Psi\,(\rho, \alpha)\big) \equiv \r{t}_\Omega\,(\TM_\Gamma\,\beta\,\rho, \r{\Gamma}\,\beta\,\alpha)
\end{equation*}

We define the presheaf $\TM^\U : \PSh\,\REN$ and a family over it
$\TM^\El : \FamPSh\,\TM^\U$. The actions on objects are $\TM^\U\,\Psi
:= \Tm\,\Psi\,\U$ and ${\TM^\El}_\Psi\,\hat{A} :=
\Tm\,\Psi\,(\El\,\hat{A})$. The action on a morphism $\beta$ is just
substitution $\blank[\cul\beta\cur]$ for both.

Note that the base category of the logical predicate interpretation is
fixed to $\REN$. However we parameterise the interpretation by the
predicate at the base type $\U$ and base family $\El$. These are
denoted by $\bar{\U}$ and $\bar{\El}$ and have the following types.
\begin{alignat*}{3}
 & \bar{\U} && : \FamPSh\,\TM^\U \\
 & \bar{\El} && : \FamPSh\,\big(\Sigma\,(\Sigma\,\TM^\U\,\TM^\El)\,\bar{\U}[\wk]\big)
\end{alignat*}

Now we list the methods for each constructor in the same order as we
have given them in section \ref{sec:object_theory}. We omit the proofs
of functoriality/naturality only for reasons of space (for these
details see \cite{kaposiphd}).

The logical predicate trivially holds at the empty context, and it
holds at an extended context for $\rho$ if it holds at the smaller
context at $\pi_1\,\rho$ and if it holds at the type which extends the
context for $\pi_2\,\rho$. The second part obviously depends on the
first. The action on morphisms for context extension is
pointwise. Here we omitted some usages of $\tr{\blank}{\blank}$
e.g.\ $\r{\Delta}\,\beta\,\alpha$ is only well-typed in that position
when we transport along the equality $\pi_1\,\rho\circ\cul\beta\cur
\equiv \pi_1\,(\rho\circ\cul\beta\cur)$. From now on we will omit
transports and the usages of $\cul\blank\cur$ in most cases for
readability.
\begin{alignat*}{4}
  & \r{\cdot}\,(\rho:\TM_\cdot\,\Psi) && := \top \\
  & \r{\cdot}\,\beta\,\_ && := \tt \\
  & \r{\Delta,A}\,(\rho:\TM_{\Delta,A}\,\Psi) && := \Sigma(\alpha : \r{\Delta}\,(\pi_1\,\rho)).\r{A}\,(\pi_1\,\rho, \pi_2\,\rho, \alpha) \\
  & \r{\Delta,A}\,(\beta : \REN(\Omega, \Psi))\,(\alpha,a) && := (\r{\Delta}\,\beta\,\alpha, \r{A}\,\beta\,a)
\end{alignat*}

The logical predicate at a substituted type is the logical predicate
at the type and we need to use the fundamental theorem at the
substitution to lift the witness of the predicate for the
substitution. Renaming a substituted type is the same as renaming in
the original type (hence the functor laws hold immediately by the
inductive hypothesis). This is well-typed because of naturality of
$\TM_\sigma$ and $\r{\sigma}$ as shown below.
\begin{alignat*}{4}
  & \r{A[\sigma]}\,(\rho, s, \alpha) && := \r{A}\,\big(\TM_\sigma\,\rho, s, \r{\sigma}\,(\rho, \alpha)\big) \\
  & \r{A[\sigma]}\,\beta\,a && := \r{A}\,\beta\,a : \underbrace{\r{A}\,\big(\TM_\Delta\,\beta\,(\TM_\sigma\,\rho), \TM_A\,\beta\,s, \r{\Delta}\,\beta\,(\r{\sigma}\,(\rho, \alpha))\big)}_{\equiv \r{A}\,\big(\TM_\sigma\,(\TM_\Gamma\,\beta\,\rho), \TM_A\,\beta\,s, \r{\sigma}\,(\TM_\Gamma\,\beta\,\rho, \r{\Gamma}\,\beta\,\alpha)\big)}
\end{alignat*}
The logical predicate at the base type and family says what we have
given as parameters. Renaming also comes from these parameters.
\begin{alignat*}{4}
  & \r{\U}\,(\rho, s, \alpha) && := \bar{\U}\,(\tr{\U[]}{s}) && \r{\U}\,\beta\,a && := \bar{\U}\,\beta\,a \\ 
  & \r{\El\,\hat{A}}\,(\rho, s, \alpha) && := \bar{\El}\,\big((\tr{\El[]}{\TM_{\hat{A}}\,\rho}), s, \r{\hat{A}}\,(\rho, \alpha)\big) \hspace{3em} && \r{\El\,\hat{A}}\,\beta\,a && := \bar{\El}\,\beta\,a 
\end{alignat*}
The logical predicate holds for a function $s$ when we have that if
the predicate holds for an argument $u$ (at $A$, witnessed by $v$), so
it holds for $s\$u$ at $B$. In addition, we have a Kripke style
generalisation: this should be true for $\TM_{\Pi\,A\,B}\,\beta\,s$
for any morphism $\beta$ in a natural way. Renaming a witness of the
logical predicate at the function type is postcomposing the Kripke
morphism by it.
\begin{alignat*}{4}
  & \r{\Pi\,A\,B}_\Psi\,\big((\rho:\TM_\Gamma\,\Psi), (s:\TM_{\Pi\,A\,B}\,\rho), (\alpha:\r{\Gamma}\,\rho)\big) \\
  & \hspace{2em} := \Sigma\Big(\map:\big(\beta:\REN(\Omega, \Psi)\big)\big(u:\TM_A\,(\TM_\Gamma\,\beta\,\rho)\big)\big(v : \r{A}_{\,\Omega}\,(\TM_\Gamma\,\beta\,\rho,u,\r{\Gamma}\,\beta\,\alpha)\big) \\
  & \hspace{7.4em} \ra \r{B}_{\,\Omega}\,\big((\TM_\Gamma\,\beta\,\rho, u), (\TM_{\Pi\,A\,B}\,\beta\,s)\$ u, (\r{\Gamma}\,\beta\,\alpha, v)\big)\Big) \\ 
  & \hspace{4.5em} . \forall\beta,u,v,\gamma . \r{B}\,\gamma\,(\map\,\beta\,u\,v) \equiv \map\,(\beta\circ\gamma)\,(\TM_A\,\gamma\,u)\,(\r{A}\,\gamma\,v) \\
  & \r{\Pi\,A\,B}\,\beta'\,(\map,\mathsf{nat}) := \lambda\beta.\map\,(\beta'\circ\beta), \lambda\beta.\mathsf{nat}\,(\beta'\circ\beta)
\end{alignat*}

Now we list the methods for the substitution constructors, that is, we
prove the fundamental theorem for substitutions. We omit the
naturality proofs. The object theoretic constructs map to their
metatheoretic counterparts: identity becomes identity, composition
becomes composition, the empty substitution becomes the element of the
unit type, substitution extension becomes pairing, first projection
becomes first projection.
\begin{alignat*}{4}
  & \r{\id}\,(\rho, \alpha) && := \alpha \\
  & \r{\sigma\circ\nu}\,(\rho, \alpha) && := \r{\sigma}\,\big(\TM_\nu\,\rho, \r{\nu}\,(\rho, \alpha)\big) \\
  & \r{\epsilon}\,(\rho, \alpha) && := \tt \\
  & \r{\sigma,t}\,(\rho, \alpha) && := \r{\sigma}\,(\rho, \alpha), \r{t}\,(\rho, \alpha) \\
  & \r{\pi_1\,\sigma}\,(\rho, \alpha) && := \proj_1\,\big(\r{\sigma}\,(\rho, \alpha)\big)
\end{alignat*}

The fundamental theorem for substituted terms and the second
projection are again just composition and second projection.
\begin{alignat*}{4}
  & \r{t[\sigma]}\,(\rho, \alpha) && := \r{t}\,(\TM_\sigma\,\rho, \r{\sigma}\,(\rho, \alpha)) \\
  & \r{\pi_2\,\sigma}\,(\rho, \alpha) && := \proj_2\,\big(\r{\sigma}\,(\rho, \alpha)\big)
\end{alignat*}
Now we prove the fundamental theorem for $\lam$ and $\app$
(i.e.\ provide the corresponding methods). For $\lam$, the $\map$
function is using the fundamental theorem for $t$ which is in the
context extended by the domain type $A : \Ty\,\Gamma$, so we need to
supply an extended substitution and a witness of the
predicate. Moreover, we need to rename the substitution $\rho$ and the
witness of the predicate $\alpha$ to account for the Kripke
property. The naturality is given by the naturality of the term
itself.
\begin{alignat*}{5}
  & \r{\lam\,t}\,(\rho, \alpha) && := \Big(\lambda\beta\,u\,v.\r{t}\,\big((\TM_\Gamma\,\beta\,\rho, u), (\r{\Gamma}\,\beta\,\alpha, v)\big) \\
  & && \hspace{1.8em} , \lambda\beta\,u\,v\,\gamma.\natS\,\r{t}\,\big((\TM_\Gamma\,\beta\,\rho, u), (\r{\Gamma}\,\beta\,\alpha, v)\big)\,\gamma\Big)
\end{alignat*}
Application uses the $\map$ part of the logical predicate and the
identity renaming.
\begin{alignat*}{5}
  & \r{\app\,t}\,(\rho, \alpha) && := \map\,\big(\r{t}\,(\pi_1\,\rho, \proj_1\,\alpha)\big)\,\id\,(\pi_2\,\rho)\,(\proj_2\,\alpha) 
\end{alignat*}

Lastly, we need to provide methods for the equality constructors. We
won't list all of these proofs as they are quite straightforward, but
as examples we show the semantic versions of the laws $[][]$ and
$\pi_2\beta$. For $[][]$, we have to show that the two families of
presheaves $\r{A[\sigma][\nu]}$ and $\r{A[\sigma\circ\nu]}$ are
equal. It is enough to show that their action on objects and morphisms
coincides as the equalities will be equal by $\mathsf{K}$. Note that
we use function extensionality to show the equality of the presheaves
from the pointwise equality of actions. When we unfold the definitions
for the actions on objects we see that the results are equal by
associativity.
\begin{alignat*}{3}
  & && \r{A[\sigma][\nu]}\,(\rho, s, \alpha) \\
  & = \, && \r{A[\sigma]}\,\big(\TM_\nu\,\rho, s, \r{\nu}\,(\rho, \alpha)\big) \\
  & = && \r{A}\,\big(\TM_\sigma\,(\TM_\nu\,\rho), s, \r{\sigma}\,(\TM_\nu\,\rho, \r{\nu}\,(\rho, \alpha))\big) \\
  & \equiv && \r{A}\,\big(\TM_{\sigma\circ\nu}\,\rho, s, \r{\sigma}\,(\TM_\nu\,\rho, \r{\nu}\,(\rho, \alpha))\big) \\
  & = && \r{A}\,\big(\TM_{\sigma\circ\nu}\,\rho, s, \r{\sigma\circ\nu}\,(\rho, \alpha)\big) \\
  & = && \r{A[\sigma\circ\nu]}\,(\rho, s, \alpha)
\end{alignat*}
The actions on morphisms are equal by unfolding the definitions.
\begin{alignat*}{6}
  & && \r{A[\sigma][\nu]}\,\beta\,a = \r{A}\,\beta\,a = \r{A[\sigma\circ\nu]}\,\beta\,a
\end{alignat*}
For $\pi_2\beta$ we need to show that two sections
$\r{\pi_2\,(\sigma,t)}$ and $\r{t}$ are equal, and again, the law
parts of the sections will be equal by $\mathsf{K}$.
\[
  {\pi_2\beta}^\M : \r{\pi_2\,(\sigma, t)}\,(\rho, \alpha) = \proj_2\,\big(\r{\sigma, t}\,(\rho, \alpha)\big) = \proj_2\,\big(\r{\sigma}\,(\rho, \alpha), \r{t}\,(\rho, \alpha)\big) = \r{t}\,(\rho, \alpha)
\]


\section{Normal forms}
\label{sec:Nf}

We define $\eta$-long $\beta$-normal forms mutually with neutral
terms. Neutral terms are terms where a variable is in a key position
which precludes the application of the rule $\Pi\beta$. Embeddings
back into the syntax are defined mutually in the obvious way. Note
that neutral terms and normal forms are indexed by types, not normal
types.
\begin{alignat*}{10}
  & \data \, \Ne && : (\Gamma:\Con) \ra \Ty\,\Gamma \ra \Set                                             && \data \, \Nf \\                                                                 
  & \data \, \Nf && : (\Gamma:\Con) \ra \Ty\,\Gamma \ra \Set                                             && \ind \neuU && : \Ne\,\Gamma\,\U \ra \Nf\,\Gamma\,\U \\                          
  & \cul \blank \cur && : \Nf\,\Gamma\,A \ra \Tm\,\Gamma\,A                                              && \ind \neuEl && : \Ne\,\Gamma\,(\El\,\hat{A}) \ra \Nf\,\Gamma\,(\El\,\hat{A}) \\ 
  & \data \, \Ne &&                                                                                      && \ind \lam && : \Nf\,(\Gamma,A)\,B \ra \Nf\,\Gamma\,(\Pi\,A\,B) \\               
  & \ind \var && : \Var\,\Gamma\,A \ra \Ne\,\Gamma\,A                                                    && \cul \blank \cur && : \Ne\,\Gamma\,A \ra \Tm\,\Gamma\,A \\
  & \ind \app && : \Ne\,\Gamma\,(\Pi\,A\,B) \ra (v : \Nf\,\Gamma\,A) \hspace{6em} \\
  & && \hspace{0.7em} \ra \Ne\,\Gamma\,(B[\lb\cul v\cur\rb])
\end{alignat*}
We define lists of neutral terms and normal forms. $X$ is a parameter
of the list, it can stand for both $\Ne$ and $\Nf$.
\begin{alignat*}{4}
  & \data\,\blank\mathsf{s} \, (X : (\Gamma : \Con) \ra \Ty\,\Gamma \ra \Set) : \Con \ra \Con \ra \Set \\
  & \cul\blank\cur : X\mathsf{s}\,\Gamma\,\Delta \ra \Tms\,\Gamma\,\Delta \\
  & \data\,X\mathsf{s} \\
  & \ind \epsilon \hspace{1.46em} : X\mathsf{s}\,\Gamma\,\cdot \\
  & \ind \blank,\blank : (\tau : X\mathsf{s}\,\Gamma\,\Delta) \ra X\,\Gamma\,A[\cul\tau\cur] \ra X\mathsf{s}\,\Gamma\,(\Delta,A)
\end{alignat*}
We also need renamings of (lists of) normal forms and neutral terms
together with lemmas relating their embeddings to terms. Again, $X$
can stand for both $\Ne$ and $\Nf$.
\begin{alignat*}{3}
  & \blank[\blank] && : X\,\Gamma\,A \ra (\beta : \Vars\,\Psi\,\Gamma) \ra X\,\Psi\,A[\cul\beta\cur] \hspace{5em} && \cul[]\cur : \cul n\cur[\cul\beta\cur] \equiv \cul n[\beta]\cur \\
  & \blank\circ\blank && : {X\mathsf{s}}\,\Gamma\,\Delta \ra \Vars\,\Psi\,\Gamma \ra X\mathsf{s}\,\Psi\,\Delta && \cul\tau\cur \circ \cul\beta\cur \equiv \cul \tau\circ\beta\cur
\end{alignat*}
Now we can define the presheaf $X_\Delta$ and families of presheaves
$X_A$ where X is either $\NE$ or $\NF$ (we use uppercase for the
families of presheaves and lowercase for the inductive types, just as
in the case of $\Tm$ and $\TM$). The definitions follow that of $\TM$.
\begin{alignat*}{6}
  & \Delta : \Con && X_\Delta && : \PSh\,\REN && X_\Delta\, \Psi && := X\mathsf{s}\,\Psi\,\Delta && X_\Delta\,\beta\,\tau := \tau\circ\beta \\
  & A : \Ty\,\Gamma \hspace{2em} && X_A && : \FamPSh\,\TM_\Gamma \hspace{2em} && X_A\,(\rho : \TM_\Gamma\,\Psi) && := X\,\Psi\,A[\rho] \hspace{2em} && X_A\,\beta\,n := n[\beta]
\end{alignat*}

We set the parameters of the logical predicate at the base type and
family by defining $\bar{\U}$ and $\bar{\El}$. The predicate holds for
a term if there is a neutral term of the corresponding type which is
equal to the term. The action on morphisms is just renaming.
\begin{alignat*}{6}
  & \bar{\U} : \FamPSh\,\TM^\U \\
  & \bar{\U}_\Psi\,(\hat{A} : \Tm\,\Psi\,\U) := \Sigma(n:\Ne\,\Psi\,\U).\hat{A}\equiv\cul n\cur \\
  & \bar{\El} : \FamPSh\,\big(\Sigma\,(\Sigma\,\TM^\U\,\TM^\El)\,\bar{\U}[\wk]\big) \\
  & \bar{\El}_\Psi\,(\hat{A}, t : \Tm\,\Psi\,(\El\,\hat{A}), p) := \Sigma(n:\Ne\,\Psi\,(\El\,\hat{A})).t\equiv\cul n\cur
\end{alignat*}

Now we can interpret any term in the logical predicate interpretation
over $\REN$ with base type interpretations $\bar{\U}$ and
$\bar{\El}$. We denote the interpretation of $t$ by $\r{t}$.

Now we turn to the proof of decidability of equality for normal
forms. Decidability of a type $X$ is defined as the sum type $\Dec\,X
:= X + (X \ra \bot)$. We start by describing a failed attempt of the
proof.

Our first try is to prove decidability directly by mutual induction
on variables, neutral terms and normal forms as given below where $X$
can be either $\Var$, $\Ne$ or $\Nf$.
\[
\dec_X : (n_0\,n_1 : X\,\Gamma\,A) \ra \Dec\,(n_0 \equiv n_1)
\]
However there is a problem with the constructor $\app$ for neutral
terms. Given $\app\,n_0\,v_0 : \Ne\,\Gamma\,(B_0[\lb\cul v_0\cur\rb])$
and $\app\,n_1\,v_1 : \Ne\,\Gamma\,(B_1[\lb\cul v_1\cur\rb])$, we
would like to first decide the equality for $n_0$ and $n_1$. If they
are not equal, we use injectivity of the neutral term constructor
$\app$ to show $\app\,n_0\,v_0 \equiv \app\,n_1\,v_1 \ra \bot$. If
they are equal, then we use the other induction hypothesis,
decidability of $v_0$ and $v_1$. However, we can't even apply the
induction hypothesis for $n_0$ and $n_1$: we would need $\Pi\,A_0\,B_0
\equiv \Pi\,A_1\,B_1$ but we only have $B_0[\lb\cul v_0\cur\rb] \equiv
B_1[\lb\cul v_1\cur\rb]$ and this doesn't give us the former
equality. It seems that we need to decide whether the types are equal
and only compare the terms afterwards (note that $A_0$ and $B_0$ are
implicit parameters of the constructor $\app$ before the $n_0$ and
$v_0$ parameters).

Hence, our next try is to generalise our induction hypothesis and
decide equality of types, not only that of terms:
\[
\dec_X : (n_0 : X\,\Gamma\,A_0)(n_1 : X\,\Gamma\,A_1) \ra \Dec\,\big(\Sigma(q : A_0 \equiv A_1).n_0 \equiv^q n_1\big)
\]
However this wouldn't work for $\lam$ because when we decide the
equality of $\lam\,v_0 : \Nf\,\Gamma\,(\Pi\,A_0\,B_0)$ and $\lam\,v_1
: \Nf\,\Gamma\,(\Pi\,A_1\,B_1)$, we would need to decide whether $v_0
: \Nf\,(\Gamma,A_0)\,B_0$ is equal to $v_1 : \Nf\,(\Gamma,A_1)\,B_1$
where the contexts are different. It seems that we need to prove
decidability of contexts, types, variables, neutral terms and normal
forms at the same time:
\[
\dec_X : (n_0 : X\,\Gamma_0\,A_0)(n_1 : X\,\Gamma_1\,A_1) \ra \Dec\,\big(\Sigma(p : \Gamma_0 \equiv \Gamma_1, q : A_0 \equiv^p A_1).n_0 \equiv^{p,q} n_1\big)
\]
However types can include non-normal terms (using $\El$), so our
current definition of normal forms seems to be not suitable for
deciding equality: we would need normal contexts and normal types
defined mutually with normal forms and neutral terms.

Before abandoning our definition of normal forms for a more
complicated one, we look at bidirectional type checking
\cite{Coquand96analgorithm}. This teaches us that for neutral terms,
the context determines the type (type inference), while for normal
forms we need a type as an input (type checking). We observe that we
can organise our induction this way: given two variables or neutral
terms of an arbitrary type in the same context, we will be able to
decide whether they are equal (including their types, which are
determined by the context). Given two normal forms of the same type
(this is the input type), we can decide whether they are equal. The
following mutual induction actually works:
\begin{alignat*}{5}
  & \dec_\Var && : (x_0 : \Var\,\Gamma\,A_0)(x_1 : \Var\,\Gamma\,A_1) \ra \Dec\,\big(\Sigma(q : A_0 \equiv A_1).x_0 \equiv^q x_1\big) \\
  & \dec_\Ne && : (n_0 : \Ne\,\Gamma\,A_0)(n_1 : \Ne\,\Gamma\,A_1) \ra \Dec\,\big(\Sigma(q : A_0 \equiv A_1).n_0 \equiv^q n_1\big) \\
  & \dec_\Nf && : (v_0\,v_1 : \Nf\,\Gamma\,A) \ra \Dec\,(v_0 \equiv v_1)
\end{alignat*}
If the variables are both $\vze$, they need to have the same type (the
last element of the context). If they are both constructed by $\vsu$,
we can use the induction hypothesis. We check whether the induction
hypothesis gives us a positive or negative result. In the positive
case, we just return the positive answer again, and in the negative
case we need to construct $x_0 \equiv x_1$ from $\vsu\,x_0 \equiv
\vsu\,x_1$ to prove $\bot$. This comes from injectivity of $\vsu$.

If the two neutral terms are variables, we use the induction
hypothesis for variables. If they are both applications, we first
decide equality of the neutral functions which will also give us an
equality of the function types. By injectivity of $\Pi$ (section
\ref{sec:inj}) we get that the domains are equal, hence we can compare
the normal arguments at this type.

If the two normal forms are neutral terms, we use the induction
hypothesis for neutral terms. If they are both $\lambda$-abstractions,
we can use the induction hypothesis thanks to injectivity of $\Pi$
again.

If two variables, neutral terms or normal forms are constructed by
different constructors, they are non equal by disjointness of
constructors.

In the above proof, we (sometimes implicitly) used injectivity of
context extension, $\El$ and $\Pi$ (section \ref{sec:inj}) and
injectivity and disjointness of the constructors for $\Var$, $\Ne$ and
$\Nf$. For all the technical details, see the formalisation
\cite{supplLMCS}.


\section{Quote and unquote}
\label{sec:quote}

By the logical predicate interpretation using $\bar{\U}$ and
$\bar{\El}$ we have the following two things:
\begin{itemize}
\item terms at the base type and base family are equal to a normal form,
\item this property is preserved by the other type formers --- this is
  what the logical predicate says at function types and substituted
  types.
\end{itemize}
We make use of this fact to lift the first property to any type. We do
this by defining a quote function by induction on the type. Quote
takes a term which preserves the predicate and maps it to a normal
form which is equal to it. Because of function spaces, we need a
function in the other direction as well, mapping neutral terms to the
witness of the predicate.

More precisely, we define the quote function $\q$ and unquote $\u$ by
induction on the structure of contexts and types. For this, we need to
define a model of type theory in which only the motives for contexts
and types are interesting.

First we define families of presheaves for contexts and types which
express that there is an equal normal form. The actions on objects are
given as follows.
\begin{alignat*}{6}
  & {\NF^\equiv}_\Delta : \FamPSh\,\TM_\Delta && {\NF^\equiv}_A : \FamPSh\,(\Sigma\,\TM_\Gamma\,\TM_A)\\
  & {\NF^\equiv}_\Delta\,(\rho : \TM_\Delta\,\Psi) := \Sigma(\rho':\NF_\Delta\,\Psi).\rho\equiv\cul\rho'\cur \hspace{2em} && {\NF^\equiv}_A\,(\rho, s) := \Sigma(s':\NF_A\,\rho).s\equiv\cul s'\cur
\end{alignat*}
We use these to write down the motives for contexts and types. We use
sections to express the commutativity of the diagram in figure
\ref{fig:diagramnew}. We only write $\Sigma$ once for iterated usage.
\begin{alignat*}{10}
  & \u_\Delta && : \NE_\Delta && \S \r{\Delta}[\cul\blank\cur] && \u_{A} && : \Sigma\,\TM_\Gamma\,\NE_A\,(\r{\Gamma}[\wk]) && \S \r{A}[\id, \cul\blank\cur, \id] \\
  & \q_\Delta && : \Sigma\,\TM_\Delta\,\r{\Delta} && \S {\NF^\equiv}_\Delta[\wk] \hspace{3em} && \q_{A} && : \Sigma\,\TM_\Gamma\,\TM_A\,(\r{\Gamma}[\wk])\,\r{A} && \S {\NF^\equiv}_A[\wk][\wk]
\end{alignat*}
Unquote for a context takes a neutral substitution and returns a proof
that the logical predicate holds for it. Quote takes a substitution
for which the predicate holds and returns a normal substitution
together with a proof that the original substitution is equal
(convertible) to the normal one (embedded into substitutions by
$\cul\blank\cur$). The types of unquote and quote for types are more
involved as they depend on a substitution for which the predicate
needs to hold. Unquote for a type takes such a substitution and a
neutral term at the type substituted by this substitution and returns
a proof that the predicate holds for this neutral term. The natural
transformation $\id, \cul\blank\cur, \id$ is defined in the obvious
way, it just embeds the second component (the neutral term) into
terms. Quote for a type takes a term of this type for which the
predicate holds and returns a normal form at this type together with a
proof that it is equal to the term. Here again, another substitution
is involved.

The motives for substitutions and terms are the constant unit
families.

We will list the methods for contexts and types omitting the
naturality proofs for brevity.

Unquote and quote for the empty context are trivial, for extended
contexts they are pointwise. ${\ap,}$ is the congruence law of
substitution extension $\blank,\blank$.
\begin{alignat*}{5}
  & \u_\cdot\,(\tau : \NE_\cdot\,\Psi) : \top := \mathsf{tt} \\
  & \q_\cdot\,\big((\sigma : \TM_\cdot\,\Psi), (\alpha : \top)\big) : \Sigma(\rho' : \NF_\cdot\,\Psi).\rho\equiv\cul\rho'\cur := (\epsilon, \epsilon\eta) \\
  & \u_{\Delta,A}\,\big((\tau, n):\NE_{\Delta,A}\,\Psi\big) : \Sigma\big(\alpha : \r{\Delta}\,(\pi_1\,\cul\tau, n\cur)\big).\r{A}\,(\pi_1\,\cul\tau, n\cur, \pi_2\,\cul\tau, n\cur, \alpha) \\
  & \hspace{1em} := \u_\Delta\,\tau, \u_A\,(\cul\tau\cur, n, \u_\Delta\,\tau) \\
  & \q_{\Delta,A}\,\big((\rho:\TM_{\Delta,A}\,\Psi), (\alpha, a) :\r{\Delta,A}\,\rho\big) : \Sigma(\rho' : \NF_{\Delta,A}\,\Psi).\rho\equiv\cul\rho'\cur \\
  & \hspace{1em} := \LET\,(\tau, p) := \q_\Delta\,(\pi_1\,\rho, \alpha);\,(n, q) := \q_A\,(\pi_1\,\rho, \pi_2\,\rho, \alpha, a) \,\IN\, \big((\tau, n), ({\ap,}\,p\,q)\big)
\end{alignat*}

(Un)quoting a substituted type is the same as (un)quoting at the type
and using the fundamental theorem at the substitution to lift the
witness of the predicate $\alpha$. As expected, unquoting at the base
type simply returns the neutral term itself and the witness of the
predicate will be reflexivity, while quote just returns the witness of
the predicate.
\begin{alignat*}{5}
  & \u_{A[\sigma]}\,(\rho, n, \alpha) && : \r{A}\,\big(\sigma\circ\rho, \cul n\cur, \r{\sigma}\,(\rho, \alpha)\big) && := \u_A\,\big(\sigma\circ\rho, n, \r{\sigma}\,(\rho, \alpha)\big) \hspace{20em} \\
  & \q_{A[\sigma]}\,(\rho, s, \alpha, a) && : \Sigma(s' : \NF_{A[\sigma]}\,\rho).s\equiv\cul s'\cur && := \q_A\,\big(\sigma\circ\rho, s, \r{\sigma}\,(\rho, \alpha), a\big)
\end{alignat*}
\begin{alignat*}{5}
  & \u_\U\,\big((\rho:\TM_\Gamma\,\Psi), (n : \Ne\,\Psi\,\U[\rho]), \alpha\big) && : \Sigma(n' : \NF_\U\,\id).\cul n\cur \equiv \cul n'\cur && := \neuU\,(\tr{\U[]}{n}), \refl \hspace{20em} \\
  & \q_\U\,\Big(\rho, t, \alpha, \big(a : {\NF^\equiv}_\U\,(\id, t)\big)\Big) && : {\NF^\equiv}_\U\,(\rho, t) && := \tr{\U[]}{a}
\end{alignat*}
\begin{alignat*}{5}
  & \u_{\El\,\hat{A}}\,\big((\rho:\TM_\Gamma\,\Psi), (n : \Ne\,\Psi\,(\El\,\hat{A}[\rho])), \alpha\big) && : \Sigma(n' : \NF_{\El\,\hat{A}}\,\id).\cul n\cur \equiv \cul n'\cur && := \neuEl\,(\tr{\El[]}{n}), \refl \hspace{20em} \\
  & \q_{\El\,\hat{A}}\,\Big(\rho, t, \alpha, \big(a : {\NF^\equiv}_{\El\,\hat{A}}\,(\id, t)\big)\Big) && : {\NF^\equiv}_{\El\,\hat{A}}\,(\rho, t) && := \tr{\El[]}{a}
\end{alignat*}
We only show the mapping part of unquoting a function. To show that
$n$ preserves the predicate, we show that it preserves the predicate
for every argument $u$ for which the predicate holds (by $v$). We
quote the argument, thereby getting it in normal form ($m$), and now
we can unquote the neutral term ($\app\,n[\beta]\,m$) to get the
result. We also need to transport the result along the proof $p$ that
$u \equiv \cul m\cur$.
\begin{alignat*}{5}
  & \map\Big(\u_{\Pi\,A\,B}\,\big((\rho:\TM_\Gamma\,\Psi), (n : \NE_{\Pi\,A\,B}\,\rho), \alpha\big)\Big)\big(\beta:\Vars\,\Omega\,\Psi\big)\big(u:\TM_A\,(\rho\circ\cul\beta\cur)\big) \\
  & \hspace{1.8em}\big(v : \r{A}_{\,\Omega}\,(\rho\circ\cul\beta\cur,u,\r{\Gamma}\,\beta\,\alpha)\big) \,\,:\,\, \r{B}_{\,\Omega}\,\big((\rho\circ\cul\beta\cur, u), (\cul n\cur[\cul\beta\cur])\$ u, (\r{\Gamma}\,\beta\,\alpha, v)\big) \\
  & \hspace{0.7em}:= \LET\,(m, p) := \q_A\,(\rho\circ\cul\beta\cur, u, \r{\Gamma}\,\beta\,\alpha, v) \, \IN \, \u_B\,\big((\rho\circ\cul\beta\cur, u), (\tr{p}{\app\,n[\beta]\,m}), (\r{\Gamma}\,\beta\,\alpha, v) \big)
\end{alignat*}
The normal form of a function $t$ is $\lam\,n$ for some normal form
$n$ which is in the extended context. We get this $n$ by quoting
$\app\,t$ in the extended context. $f$ is the witness that $t$
preserves the relation for any renaming, and we use the renaming
$\wkV\,\id$ to use $f$ in the extended context. The argument of $f$ in
this case will be the zero de Bruijn index $\vze$ and we need to
unquote it to get the witness that it preserves the logical
predicate. This is the place where the Kripke property of the logical
relation is needed: the base category of the Kripke logical relation
needs to minimally include the morphism $\wkV\,\id$ (in our case it
has type $\Vars\,(\Gamma,A)\,\Gamma$).
\begin{alignat*}{5}
  & \q_{\Pi\,A\,B}\,(\rho, t, \alpha, f) : \Sigma(t' : \NF_{\Pi\,A\,B}\,\rho).t\equiv\cul t'\cur \\
  & \hspace{1em} := \LET\, a \hspace{1.8em} := \u_A\,(\rho\circ\cul\wkV\,\id\cur, \var\,\vze, \r{\Gamma}\,(\wkV\,\id)\,\alpha) \\
  & \hspace{3.8em} (n, p) := \q_B\,\big(\rho^A, \app\,t, (\r{\Gamma}\,(\wkV\,\id)\,\alpha, a), \map\,f\,(\wkV\,\id)\,\cul\vze\cur\,a \big) \\
  & \hspace{2.6em} \IN\hspace{0.5em} (\lam\,n, {\Pi\eta}^{-1} \trans \ap\,\lam\,p)
\end{alignat*}
We have to verify the equality laws for types. Note that we use
function extensionality to show that the corresponding quote and
unquote functions are equal. The naturality proofs will be equal by
$\mathsf{K}$.

(Un)quote preserves $[\id]$ by the left identity law.
\begin{alignat*}{5}
  & \u_{A[\id]}\,(\rho, n, \alpha) && = \u_A\,(\id\circ\rho, n, \alpha) && \equiv \u_{A}\,(\rho, n, \alpha) \\
  & \q_{A[\id]}\,(\rho, s, \alpha, a) && = \q_A\,(\id\circ\rho, s, \alpha, a) && \equiv \q_{A}\,(\rho, s, \alpha, a)
\end{alignat*}
(Un)quote preserves $[][]$ by associativity for substitutions.
\begin{alignat*}{5}
  & && \u_{A[\sigma][\nu]}\,(\rho, n, \alpha) \\
  & =\, && \u_{A}\,\big(\sigma\circ(\nu\circ\rho), n, \r{\sigma}\,(\nu\circ\rho, \r{\nu}\,(\rho, \alpha))\big) \\
  & \equiv && \u_A\,\big((\sigma\circ\nu)\circ\rho, n, \r{\sigma}\,(\nu\circ\rho, \r{\nu}\,(\rho, \alpha))\big) \\
  & = && \u_{A[\sigma\circ\nu]}\,(\rho, n, \alpha)
\end{alignat*}
\begin{alignat*}{5}
  & && \q_{A[\sigma][\nu]}\,(\rho, s, \alpha, a) \\
  & =\, && \q_A\,\big(\sigma\circ(\nu\circ\rho), s, \r{\sigma}\,(\nu\circ\rho, \r{\nu}\,(\rho, \alpha)), a\big) \\
  & \equiv && \q_A\,((\sigma\circ\nu)\circ\rho, s, \r{\sigma}\,(\nu\circ\rho, \r{\nu}\,(\rho, \alpha)), a) \\
  & = && \q_{A[\sigma\circ\nu]}\,(\rho, s, \alpha, a)
\end{alignat*}
The semantic counterparts of $\U[]$ and $\El[]$ are verified as
follows.
\begin{alignat*}{5}
  & && \u_{\U[\sigma]}\,(\rho, n, \alpha) && = \u_\U\,\big(\sigma\circ\rho, n, \r{\sigma}\,(\rho, \alpha)\big) && = (n, \refl) && = \u_{\U}\,(\rho, n, \alpha) \\
  & && \q_{\U[\sigma]}\,(\rho, t, \alpha, a) && = \q_\U\,\big(\sigma\circ\rho, t, \r{\sigma}\,(\rho, \alpha), a\big) && = a && = \q_{\U}\,(\rho, t, \alpha, a) \\
  & && \u_{(\El\,\hat{A})[\sigma]}\,(\rho, n, \alpha) && = \u_{\El\,\hat{A}}\,\big(\sigma\circ\rho, n, \r{\sigma}\,(\rho, \alpha)\big) && = (n, \refl) && =\, \u_{\El\,(\hat{A}[\sigma])}\,(\rho, n, \alpha) \\
  & && \q_{(\El\,\hat{A})[\sigma]}\,(\rho, t, \alpha, a) && = \q_{\El\,\hat{A}}\,\big(\sigma\circ\rho, t, \r{\sigma}\,(\rho, \alpha), a\big) && = a && = \q_{\El\,(\hat{A}[\sigma])}\,(\rho, t, \alpha, a)
\end{alignat*}
For reasons of space, we only state what we need to verify for
$\Pi[]$. It is enough to show that the mapping parts of the unquoted
functions are equal and that the first components of the results of
quote are equal because the other parts are equalities.
\begin{alignat*}{5}
  & \map\,\big(\u_{(\Pi\,A\,B)[\sigma]}\,(\rho, n, \alpha)\big) && \equiv \map\,\big(\u_{\Pi\,A[\sigma]\,B[\sigma\uparrow A]}\,(\rho, n, \alpha)\big) \\
  & \proj_1\,\big(\q_{(\Pi\,A\,B)[\sigma]}\,(\rho, t, \alpha, f)\big) && \equiv \proj_1\,\big(\q_{\Pi\,A[\sigma]\,B[\sigma\uparrow A]}\,(\rho, t, \alpha, f)\big)
\end{alignat*}

The methods for substitutions and terms (including the equality
methods) are all trivial.


\section{Reaping the fruits}
\label{sec:fruits}

Now we can define the normalisation function and show that it is
complete as follows. We quote at the type of the input term and as
parameters we provide the identity substitution, the term itself, a
witness that the logical predicate holds for the identity (neutral)
substitution (this is given by unquote) and a witness that the
predicate holds for the input term. This is given by the fundamental
theorem of the logical relation, which needs identity again.
\begin{alignat*}{4}
  & \norm_A\,&& (t : \Tm\,\Gamma\,A) && : \Nf\,\Gamma\,A && := \proj_1\,\Big(\q_A\,\big(\id, t, \u_\Gamma\,\id, \r{t}\,(\id, \u_\Gamma\,\id) \big)\Big) \\
  & \mathsf{compl}_A\,&& (t : \Tm\,\Gamma\,A) && : t \equiv \cul\norm_A\,t\cur && := \proj_2\,\Big(\q_A\,\big(\id, t, \u_\Gamma\,\id, \r{t}\,(\id, \u_\Gamma\,\id) \big)\Big)
\end{alignat*}
Note that we implicitly used the equality $A[\id]\equiv A$ in the
above definitions.

We prove stability by mutual induction on neutral terms and normal
forms.
\begin{alignat*}{5}
  & \stab_\Ne && : (n : \Ne\,\Gamma\,A) \ra \r{\cul n\cur}\,(\id, \u_\Gamma\,\id) \equiv \u_A\,(\id, n, \u_\Gamma\,\id) \\
  & \stab_\Nf && : (n : \Nf\,\Gamma\,A) \ra \norm_A\,\cul n \cur\equiv n
\end{alignat*}
Decidability of equality comes from that for normal forms.
\begin{alignat*}{5}
  & \dec\,(t_0\,t_1 : \Tm\,\Gamma\,A) : \Dec\,(t_0 \equiv t_1) := \tr{\mathsf{compl}_A\,t_1}{\big(\tr{\mathsf{compl}_A\,t_0}{\dec_\Nf\,(\norm_A\,t_0)\,(\norm_A\,t_1)}\big)}
\end{alignat*}
Similarly, consistency of normal forms can be proven by the following
mutual induction.
\[
\cons_\Var : \Var\,\cdot\,A\ra \bot \hspace{5em} \cons_\Ne : \Ne\,\cdot\,A\ra \bot \hspace{5em} \cons_\Nf : \Nf\,\cdot\,\U \ra \bot
\]
It follows that our theory is consistent.
\[
\cons\,(t : \Tm\,\cdot\,\U) : \bot := \tr{\mathsf{compl}_\U\,t}{\cons_\Nf\,(\norm_A\,t)}
\]


\section{Conclusions and further work}
\label{sec:discussion}

We proved normalisation for a basic type theory with dependent types
by the technique of NBE. We evaluate terms into a proof relevant
logical predicate model. The model is depending on the syntax, we need
to use the dependent eliminator of the syntax. Our approach can be
seen as merging the presheaf model and the logical relation used in
NBE for simple types \cite{alti:ctcs95} into a single logical
predicate. This seems to be necessary because of the combination of
type indexing and dependent types: the well-typedness of normalisation
depends on completeness. Another property to note is that we don't
normalise types, we just index normal terms by not necessarily normal
types.

QIITs make it possible to define the syntax of type theory in a very
concise way, however because of missing computation rules, reasoning
with them involves lots of boilerplate. We expect that a cubical
metatheory \cite{ctt} with its systematic way of expressing equalities
depending on equalities and its additional computation rules would
significantly reduce the amount of boilerplate. The metatheoretic
status of QIITs is not yet clear, however we hope that we can justify
them using a setoid model \cite{alti:lics99}. In addition, work on the
more general higher inductive types would also validate these
constructions.

Another challenge is to extend our basic type theory with inductive
types, universes and large elimination. Also, it would be interesting
to see how the work fits into the setting of homotopy type theory
(without assuming $\mathsf{K}$). We would also like to investigate
whether the logical predicate interpretation can be generalised to
work over arbitrary presheaf models and study its relation to
categorical glueing.


\section*{Acknowledgements}
We would like to thank Bernhard Reus and the anonymous reviewers for
their helpful comments and suggestions.


\bibliographystyle{plain}
\bibliography{local,alti}{}

\begin{thebibliography}{10}

\bibitem{agdawiki}
{The Agda development team}.
\newblock The {Agda} {Wiki}, 2017.
\newblock Available online.

\bibitem{abel2010towards}
Andreas Abel.
\newblock Towards normalization by evaluation for the $\beta$$\eta$-calculus of
  constructions.
\newblock In {\em Functional and Logic Programming}, pages 224--239. Springer,
  2010.

\bibitem{abel2013normalization}
Andreas Abel.
\newblock {\em Normalization by Evaluation: Dependent Types and
  Impredicativity}.
\newblock PhD thesis, Habilitation, Ludwig-Maximilians-Universit{\"a}t
  M{\"u}nchen, 2013.

\bibitem{abel2007normalization}
Andreas Abel, Thierry Coquand, and Peter Dybjer.
\newblock Normalization by evaluation for {Martin-L\"of} type theory with typed
  equality judgements.
\newblock In {\em Logic in Computer Science, 2007. LICS 2007. 22nd Annual IEEE
  Symposium on}, pages 3--12. IEEE, 2007.

\bibitem{abelscherer}
Andreas Abel and Gabriel Scherer.
\newblock On irrelevance and algorithmic equality in predicative type theory.
\newblock {\em Logical Methods in Computer Science}, 8(1), 2012.

\bibitem{alti:lics99}
Thorsten Altenkirch.
\newblock Extensional equality in intensional type theory.
\newblock In {\em 14th Symposium on Logic in Computer Science}, pages 412 --
  420, 1999.

\bibitem{alti:lics01}
Thorsten Altenkirch, Peter Dybjer, Martin Hofmann, and Phil Scott.
\newblock Normalization by evaluation for typed lambda calculus with
  coproducts.
\newblock In {\em 16th Annual IEEE Symposium on Logic in Computer Science},
  pages 303--310, 2001.

\bibitem{alti:ctcs95}
Thorsten Altenkirch, Martin Hofmann, and Thomas Streicher.
\newblock Categorical reconstruction of a reduction free normalization proof.
\newblock In David Pitt, David~E. Rydeheard, and Peter Johnstone, editors, {\em
  Category Theory and Computer Science}, LNCS 953, pages 182--199, 1995.

\bibitem{alti:lics96}
Thorsten Altenkirch, Martin Hofmann, and Thomas Streicher.
\newblock Reduction-free normalisation for a polymorphic system.
\newblock In {\em 11th Annual IEEE Symposium on Logic in Computer Science},
  pages 98--106, 1996.

\bibitem{alti:f97}
Thorsten Altenkirch, Martin Hofmann, and Thomas Streicher.
\newblock Reduction-free normalisation for system {$F$}.
\newblock 1997.

\bibitem{supplLMCS}
Thorsten Altenkirch and Ambrus Kaposi.
\newblock Agda formalisation for the paper {Normalisation by Evaluation for
  Type Theory, in Type Theory}, 2016.
\newblock Available online at the second author's website.

\bibitem{nbe-conf}
Thorsten Altenkirch and Ambrus Kaposi.
\newblock Normalisation by evaluation for dependent types.
\newblock In {\em 1st International Conference on Formal Structures for
  Computation and Deduction, {FSCD} 2016, June 22-26, 2016, Porto, Portugal},
  pages 6:1--6:16, 2016.

\bibitem{ttintt}
Thorsten Altenkirch and Ambrus Kaposi.
\newblock Type theory in type theory using quotient inductive types.
\newblock In {\em Proceedings of the 43rd Annual ACM SIGPLAN-SIGACT Symposium
  on Principles of Programming Languages}, POPL 2016, pages 18--29, New York,
  NY, USA, 2016. ACM.

\bibitem{berger1991inverse}
Ulrich Berger and Helmut Schwichtenberg.
\newblock An inverse of the evaluation functional for typed $\lambda$-calculus.
\newblock In {\em Logic in Computer Science, 1991. LICS'91., Proceedings of
  Sixth Annual IEEE Symposium on}, pages 203--211. IEEE, 1991.

\bibitem{cockxsprinkles}
Jesper Cockx and Andreas Abel.
\newblock Sprinkles of extensionality for your vanilla type theory.
\newblock In Silvia Ghilezan and Ivetić Jelena, editors, {\em 22nd
  International Conference on Types for Proofs and Programs, TYPES 2016}.
  University of Novi Sad, 2016.

\bibitem{ctt}
Cyril Cohen, Thierry Coquand, Simon Huber, and Anders M{\"o}rtberg.
\newblock Cubical type theory: a constructive interpretation of the univalence
  axiom.
\newblock December 2015.

\bibitem{Coquand96analgorithm}
Thierry Coquand.
\newblock An algorithm for type-checking dependent types.
\newblock {\em Science of Computer Programming}, 26:167--177, 1996.

\bibitem{crole}
Roy~L. Crole.
\newblock {\em Categories for types}.
\newblock Cambridge mathematical textbooks. Cambridge University Press,
  Cambridge, New York, 1993.

\bibitem{danielsson2006formalisation}
Nils~Anders Danielsson.
\newblock A formalisation of a dependently typed language as an
  inductive-recursive family.
\newblock In {\em Types for Proofs and Programs}, pages 93--109. Springer,
  2006.

\bibitem{danvy1999type}
Olivier Danvy.
\newblock {\em Type-directed partial evaluation}.
\newblock Springer, 1999.

\bibitem{dybjer1996internal}
Peter Dybjer.
\newblock Internal type theory.
\newblock In {\em Types for Proofs and Programs}, pages 120--134. Springer,
  1996.

\bibitem{harperpfenning}
Robert Harper and Frank Pfenning.
\newblock On equivalence and canonical forms in the lf type theory.
\newblock {\em ACM Trans. Comput. Logic}, 6(1):61--101, January 2005.

\bibitem{hofmann95conservativity}
Martin Hofmann.
\newblock Conservativity of equality reflection over intensional type theory.
\newblock In {\em TYPES 95}, pages 153--164, 1995.

\bibitem{hofmann1997syntax}
Martin Hofmann.
\newblock Syntax and semantics of dependent types.
\newblock In {\em Extensional Constructs in Intensional Type Theory}, pages
  13--54. Springer, 1997.

\bibitem{kaposiphd}
Ambrus Kaposi.
\newblock {\em Type theory in a type theory with quotient inductive types}.
\newblock PhD thesis, University of Nottingham, 2016.

\bibitem{forsberg-phd}
Fredrik Nordvall~Forsberg.
\newblock {\em Inductive-inductive definitions}.
\newblock PhD thesis, Swansea University, 2013.

\bibitem{agda}
Ulf Norell.
\newblock {\em Towards a practical programming language based on dependent type
  theory}.
\newblock PhD thesis, Chalmers University of Technology, 2007.

\bibitem{Oury2005}
Nicolas Oury.
\newblock {\em Extensionality in the calculus of constructions}, pages
  278--293.
\newblock Springer Berlin Heidelberg, Berlin, Heidelberg, 2005.

\bibitem{book}
{The Univalent Foundations Program}.
\newblock Homotopy type theory: Univalent foundations of mathematics.
\newblock Technical report, Institute for Advanced Study, 2013.

\end{thebibliography}

\appendix
\section{Categorical definitions}
\label{appendix}

We use the following categorical definitions in sections
\ref{sec:logpred}, \ref{sec:Nf} and \ref{sec:quote}. Note that we work
in a setting with $\mathsf{K}$ (uniqueness of identity proofs).

A category $\C$ is given by a type of objects $|\C|$ and given $I, J :
|\C|$, a type $\C(I, J)$ which we call the type of morphisms between
$I$ and $J$. A category is equipped with an operation for composing
morphisms $\blank\circ\blank : \C(J, K) \ra \C(I, J) \ra \C(I, K)$ and
an identity morphism at each object $\id_I : \C(I, I)$. In addition we
have the associativity law $(f\circ g)\circ h \equiv f\circ(g\circ h)$
and the identity laws $\id\circ f \equiv f$ and $f\circ\id \equiv
f$.

A contravariant presheaf over a category $\C$ is denoted
$\Gamma:\PSh\,\C$. It is given by the following data: given $I :
|\C|$, a set $\Gamma\,I$, and given $f : \C(J,I)$ a function
$\Gamma\,f : \Gamma\,I \ra \Gamma\,J$. Moreover, we have $\idP\,\Gamma
: \Gamma\,\id\,\alpha \equiv \alpha$ and $\compP\,\Gamma : \Gamma\,(f
\circ g)\,\alpha \equiv \Gamma\,g\, (\Gamma\,f\,\alpha)$ for $\alpha :
\Gamma\,I$, $f:\C(J,I)$, $g:\C(K,J)$.

Given $\Gamma : \PSh\,\C$, a family of presheaves over $\Gamma$ is
denoted $A : \FamPSh\,\Gamma$. It is given by the following data
(indeed, this is equivalent to a presheaf over the category of
elements $\int \Gamma$) : given $\alpha : \Gamma\,I$, a set
$A_I\,\alpha$ and given $f : \C(J,I)$, a function $A\,f : A_I\,\alpha
\ra A_J\,(\Gamma\,f\,\alpha)$. In addition, we have the functor laws
$\idF\,A:A\,\id\,v \equiv^\idP v$ and $\compF\,A:A\,(f\circ g)\,v
\equiv^\compP A\,g\,(A\,f\,v)$ for $\alpha : \Gamma\,I$, $v :
A_I\,\alpha$, $f:\C(J,I)$, $g:\C(K,J)$.

A natural transformation between presheaves $\Gamma$ and $\Delta$ is
denoted $\sigma : \Gamma \nat \Delta$. It is given by a function
$\sigma : \{I : |\C|\} \ra \Gamma\,I \ra \Delta\,I$ together with the
condition $\natn\,\sigma:\Delta\,f\,(\sigma_I\,\alpha) \equiv
\sigma_J\,(\Gamma\,f\,\alpha)$ for $\alpha : \Gamma\,I$, $f :
\C(J,I)$.

A section from a presheaf $\Gamma$ to a family of presheaves $A$ over
$\Gamma$ is denoted $t : \Gamma \S A$. It is given by a function $t :
\{I : |\C|\}\ra(\alpha : \Gamma\,I) \ra A_I\,\alpha$ together with the
naturality condition $\natS\,t\,\alpha\,f : A\,f\,(t\,\alpha) \equiv
t\,(\Gamma\,f\,\alpha)$ for $f : \C(J, I)$. We call this a section as
it can be viewed as a section of the first projection from
$\Sigma\,\Gamma\,A$ to $\Gamma$ but we define it without using the
projection.

Given $\Gamma:\PSh\,\C$ and $A : \FamPSh\,\Gamma$ we can define
$\Sigma\,\Gamma\,A : \PSh\,\C$ by $(\Sigma\,\Gamma\,A)\,I :=
\Sigma(\alpha:\Gamma\,I).A_I\,\alpha$ and
$(\Sigma\,\Gamma\,A)\,f\,(\alpha,a):=(\Gamma\,f\,\alpha, A\,f\,a)$.

Given $\sigma : \Gamma \nat \Delta$ and $A : \FamPSh\,\Delta$, we
define $A[\sigma]:\FamPSh\,\Gamma$ by $A[\sigma]_I\,\alpha :=
A_I\,(\sigma_I\,\alpha)$ and $A[\sigma]\,f\,a :=
\tr{\natn\,\sigma}{(A\,f\,a)}$ for $\alpha:\Gamma\,I$, $a :
A[\sigma]\,\alpha$ and $f:\C(J,I)$.

The weakening natural transformation $\wk : \Sigma\,\Gamma\,A \nat
\Gamma$ is defined by $\wk_I\,(\alpha,a) := \alpha$.

Lifting of a section $t : \Gamma \S A$ by a family of presheaves $B :
\FamPSh\,\Gamma$ is a natural transformation $t\uparrow B :
\Sigma\,\Gamma\,B \nat \Sigma\,(\Sigma\,(\Gamma\,A))\,B[\wk]$. It is
defined as $(t\uparrow B)_I\,(\alpha,b) := (\alpha, t_I\,\alpha, b)$.

\end{document}